\documentclass[a4paper,10pt]{article}
\pdfoutput=1 

\usepackage{jheppub} 
\usepackage{soul}
\usepackage[T1]{fontenc} 

\usepackage{amsfonts}
\usepackage{amsmath}
\usepackage{amsthm}
\usepackage{amssymb}
\usepackage{array}
\usepackage{dcolumn}
\usepackage{placeins}
\usepackage[svgnames]{xcolor}
\usepackage{graphics}
\usepackage{tikz}
\usepackage{graphicx}
\usetikzlibrary{positioning}
\usepackage{caption}
\usepackage{subcaption}
\usepackage{adjustbox}
\usepackage{gensymb}
\usepackage{breqn}
\usepackage{braket}
\usepackage{enumitem}
\usepackage{makecell}
\usepackage{ragged2e}
\usepackage{comment}
\usepackage{url}
\usepackage{placeins}
\usepackage{blindtext}
\usepackage{float}
\usepackage{hyperref}
\usepackage{fontawesome}
\usepackage{cleveref}
\usepackage{upgreek}
\crefname{section}{Sec.}{Secs.}
\crefname{figure}{Fig.}{Figs.}
\crefname{equation}{Eq.}{Eqs.}
\crefname{appendix}{Appendix}{Appendices}

\preprint{MI-HET-823, KIAS-P23069}

\newcommand\be{\begin{equation}}
\newcommand\ee{\end{equation}}
\newcommand\bea{\begin{eqnarray}}
\newcommand\eea{\end{eqnarray}}

\definecolor{nicegreen}{rgb}{0., 0.75, 0.46}

\crefname{section}{Sec.}{Secs.}
\crefname{figure}{Fig.}{Figs.}
\crefname{equation}{Eq.}{Eqs.}
\crefname{table}{Table}{Tables}

\title{\boldmath Non-standard neutrino interactions mediated by a light scalar at DUNE}

\author[a]{Bhaskar Dutta,}
\author[b]{Sumit Ghosh,}
\author[a]{Kevin J. Kelly}
\author[c, d]{Tianjun Li,}
\author[a,e]{Adrian Thompson,}
\author[a]{and Ankur Verma}

\affiliation[a]{Mitchell Institute for Fundamental Physics and Astronomy, Department of Physics  and Astronomy, Texas A$\&$M University, College Station, Texas 77843,  USA}

\affiliation[b]{School of Physics, Korea Institute for Advanced Study, Seoul 02455, Korea}

\affiliation[c]{CAS Key Laboratory of Theoretical Physics, Institute of Theoretical Physics,
Chinese Academy of Sciences, Beijing, 100190, People’s Republic China}

\affiliation[d]{School of Physical Sciences, University of Chinese Academy of Sciences,
Beijing 100049, People’s Republic China}

\affiliation[e]{Department of Physics and Astronomy, Northwestern~University,~Evanston,~IL~60208,~USA}

\emailAdd{dutta@physics.tamu.edu}
\emailAdd{ghosh@kias.re.kr}
\emailAdd{kjkelly@tamu.edu}
\emailAdd{tli@itp.ac.cn}
\emailAdd{thompson@tamu.edu}
\emailAdd{averma1@tamu.edu}

\abstract{We investigate the effect on neutrino oscillations generated by beyond-the-standard-model interactions between neutrinos and matter. Specifically, we focus on scalar-mediated non-standard interactions (NSI) whose impact fundamentally differs from that of vector-mediated NSI. Scalar NSI contribute as corrections to the neutrino mass matrix rather than the matter potential and thereby predict distinct phenomenology from the vector-mediated ones. Similar to vector-type NSI, the presence of scalar-mediated neutrino NSI can influence measurements of oscillation parameters in long-baseline neutrino oscillation experiments, with a notable impact on CP measurement in the case of DUNE. Our study focuses on the effect of scalar NSI on neutrino oscillations, using DUNE as an example.  We introduce a model-independent parameterization procedure that enables the examination of the impact of all non-zero scalar NSI parameters simultaneously. Subsequently, we convert DUNE's sensitivity to the NSI parameters into projected sensitivity concerning the parameters of a light scalar model. We compare these results with existing non-oscillation probes. Our findings reveal that the region of the light scalar parameter space sensitive to DUNE is predominantly excluded by non-oscillation probes, except for scenarios with very light mediator mass.}  


\begin{document} 
\maketitle
\flushbottom

\section{Introduction}

The discovery of neutrino oscillations established that physics beyond the Standard Model (SM) exists in the phenomenon of neutrino oscillations~\cite{Mikheev:1986gs, Smirnov:2004zv, Wolfenstein:1977ue, SNO:2002tuh, KamLAND:2002uet, T2K:2011ypd, DoubleChooz:2011ymz, DayaBay:2012fng, RENO:2012mkc}. Since the earliest studies of neutrino oscillations, neutrino interactions with SM matter and their impact on oscillations have been a prime focus. Specifically, neutrino-electron interactions in the Sun lead to the MSW effect~\cite{Smirnov:2004zv, Mikheev:1986gs}, and neutrino-matter interactions in the Earth modify oscillation probabilities significantly for neutrinos between their production and detection in experiments such as T2K~\cite{T2K:2011qtm}, NOvA~\cite{NOvA:2007rmc}, IceCube~\cite{IceCube:2006tjp}, and Super-Kamiokande~\cite{Super-Kamiokande:2004orf}. Such effects are all predictions of the simplest extensions of the SM to account for neutrino masses and do not require any further beyond-the-SM (BSM) neutrino interactions.

The additional BSM neutrino interactions have been studied in great detail under the moniker of neutrino non-standard interactions (NSIs) that posit additional neutrino-SM (typically electrons and quarks) interactions that can be detectable in laboratory environments. This includes modifications to neutrino oscillations incurred by non-standard matter effects~\cite{Gavela:2008ra,Antusch:2008tz,Ohlsson:2012kf,Proceedings:2019qno} as well as more direct searches for modified neutrino-target scattering rates and spectra, e.g., in coherent elastic neutrino-nucleus (CEvNS) scattering~\cite{AristizabalSierra:2018eqm,Giunti:2019xpr}.

Typically, such NSIs are assumed to be mediated by vector/axial-vector interactions \`{a} la those from the SM $W$- and $Z$-bosons for neutrinos. In this work, we focus instead on scalar non-standard interactions (SNSIs), where a new scalar particle is introduced to mediate interactions between the SM particles. Light and weakly-coupled scalar-mediated NSIs between neutrinos and SM fermions can naturally emerge in numerous UV-complete models at low energy scales (see, for example Ref.~\cite{Dutta:2022fdt}). Various laboratory-based experiments involving solar and reactor neutrino scattering, as well as cosmological and astrophysical probes, enable searches in the parameter space associated with light scalar fields. The existence of these new interactions can also be explored through neutrino oscillation experiments.

The presence of SNSI manifests as a correction to the neutrino mass matrix in the evolution Hamiltonian. Consequently, the BSM modifications to neutrino oscillation probabilities follow a different structure (in terms of scaling with matter densities, baseline lengths, and neutrino energies) than that predicted by vector-mediated NSI which modify the matter potential in the time-evolution Hamiltonian. Given that the impact of scalar NSI scales linearly with the matter density of the medium, long-baseline experiments such as Deep Underground Neutrino Experiment (DUNE)~\cite{DUNE:2020ypp}, JUNO~\cite{JUNO:2015zny}, T2K~\cite{T2K:2011qtm}, and NOvA~\cite{NOvA:2007rmc} provide an ideal environment for probing these effects. In particular, it can mimic the CP measurement efforts in oscillation experiments and prevent the accurate measurement of the leptonic Dirac CP phase angle. Recently, such possibilities have been extensively explored by formulating SNSI in a manner that primarily allows for the exploration of one parameter at a time~\cite{Ge:2018uhz, Ge:2019tdi, Denton:2022pxt, Medhi:2021wxj, Medhi:2022qmu, Medhi:2023ebi, Singha:2023set, Sarker:2023qzp, Gupta:2023wct, ESSnuSB:2023lbg}.

In this work, we introduce a new method for studying SNSI in neutrino oscillations, focusing on a parameterization that allows for more than one nonzero SNSI parameters to be studied at a time, while focusing on the direct impact on the neutrino oscillation Hamiltonian. We discuss how constraints and sensitivities using this new phenomenological approach may be compared against those depending on the underlying model parameters, i.e., Yukawa couplings and the mass of the new particle. We use the upcoming DUNE experiment as a test case in understanding this dynamic, as well as the mild T2K/NOvA tension~\cite{T2K:2021xwb,T2K:2023smv,NOvA:2021nfi,NOvA:2023iam,Kelly:2020fkv,Esteban:2020cvm,deSalas:2020pgw} that has been highlighted as potential motivation for SNSI in recent studies~\cite{Denton:2022pxt}.

The remainder of this work is organized as follows. First, in section~\ref{sec:scalarNSI}, we introduce the SNSI scenario under study here, review how such SNSI impact neutrino oscillations, and introduce the phenomenological parameterization for studies of this type. In section~\ref{sec:DUNE}, we demonstrate the DUNE sensitivity in this approach and discuss how these constraints may be compared against other results in the literature. Section~\ref{Section:constraints} then performs a broad comparison of existing experimental constraints on the SNSI parameter space, while our results are then shown in section~\ref{sec:results}. Finally in Section~\ref{sec:conclusion} we offer some remarks on the future study of SNSI. Additional details concerning the numerical results are provided in the appendix. 

\section{Scalar NSI Formalism for Oscillations} \label{sec:scalarNSI}
In the standard three-neutrino mixing framework, neutrino oscillations are described by a Hamiltonian for (unitary) time-evolution through a medium. This Hamiltonian, along with the misalignment between neutrino mass and flavor eigenstates (described by the leptonic mixing matrix $U$), determines transition amplitudes and oscillation probabilities for neutrinos traveling over some distance. The standard picture also allows for neutrino interactions with matter induced by the SM $W-$ and $Z-$ boson exchange.

Such SM interactions, as well as vector-like non-standard interactions (NSI)~\cite{Proceedings:2019qno} add an effective potential to the propagation Hamiltonian, e.g.,
\begin{equation}
H_{ij} = H_{ij}^0 + U^\dagger V_{\alpha\beta} U,
\end{equation}
where $H_{ij}^0$ is the Hamiltonian for neutrino evolution in vacuum in the neutrino-mass basis and $V_{\alpha\beta}$ describes the vector-like interactions between neutrinos and SM matter during propagation. In the SM, $V_{\alpha\beta} = \mathrm{diag}(V_{\rm CC}, 0,0)$, $V_{\rm CC} = \sqrt{2} G_F n_e$ ($G_F$ being the Fermi constant and $n_e$ being the number density of electrons along the path).\footnote{Despite SM interactions with protons and neutrons existing, they make no impact on the propagation Hamiltonian because such interactions provide an overall diagonal contribution to $V_{\alpha\beta}$ to which neutrino oscillations are insensitive.} Additional BSM vector-like NSIs amount to modifying $V_{\alpha\beta}$, including the possibility of inducing off-diagonal, complex contributions that lead to additional CP violation in neutrino oscillations. We refer the reader to Refs.~\cite{deGouvea:2015ndi,Coloma:2015kiu} for further detail on vector-like NSI in the context of long-baseline oscillation experiments such as DUNE.

In this work, we focus on the less-explored possibility that NSIs are generated by scalar mediators. Scalar NSIs can be described by the following Lagrangian that includes the Yukawa interaction terms for Dirac neutrinos and SM fermions $f$
\begin{equation}
\mathcal{L}_{\phi^{\prime}}=\bar{\nu}\left(i \gamma^\mu \partial_\mu-m_\nu\right) \nu-(y_\nu)_{\alpha \beta} \bar{\nu}_{\alpha} \nu_{\beta} \phi^{\prime}-\left( y_f\right)_{\alpha \beta} \bar{f}_\alpha f_\beta \phi^{\prime}-\frac{1}{2}\left(\partial_\mu \phi^{\prime}\right)^2-\frac{m_{\phi^{\prime}}^2}{2} \phi^{\prime 2}
\end{equation}
The Yukawa interaction terms, responsible for the generation of the SNSI, can arise naturally in a UV-complete model with an extended scalar sector. Here $\phi^\prime$ is a light real scalar field that emerges from the CP-even Higgs sector (see, for example, Refs.~\cite{Dutta:2020scq, Dutta:2022fdt}). As we discussed above, vector-like NSIs modify $V_{\alpha\beta}$, where scalar interactions will generate an effective mass for the neutrinos, e.g.,
\begin{equation}
    \mathbb{M} \to \mathbb{M}_0 + \delta\mathbb{M}_{ij}
\end{equation}
We have implicitly assumed here that $\mathbb{M}_0$ (the standard, diagonal mass matrix) and $\delta\mathbb{M}_{ij}$ are written in the neutrino mass basis. However in the literature, the corrections to the mass terms are usually parameterized in the flavor basis as follows
\begin{equation}\label{eq:deltaMalphabeta}
    \delta \mathbb{M}_{\alpha\beta} \equiv \sqrt{|\Delta m_{31}^2|} \begin{pmatrix}
 \eta_{ee}  & \eta_{e\mu}e^{i \phi_{e\mu}} & \eta_{e\tau}e^{i\phi_{e\tau}}\\
\eta_{e\mu}e^{-i\phi_{e\mu}} &  \eta_{\mu\mu}  & \eta_{\mu\tau}e^{i \phi_{\mu\tau}} \\
\eta_{e\tau} e^{-i \phi_{e\tau}} & \eta_{\mu\tau} e^{-i \phi_{\mu\tau}} & \eta_{\tau\tau} 
\end{pmatrix},
\end{equation}
such that $\delta \mathbb{M}_{ij} \equiv (U^\dagger)_{i\alpha} \delta \mathbb{M}_{\alpha\beta} U_{\beta j}$. Here the elements of $\delta \mathbb{M}_{\alpha\beta}$ are expressed relative to $\sqrt{|\Delta m_{31}^2|}$ in order to ensure that $\eta_{\alpha \beta}$ are dimensionless. Hermiticity in the neutrino Hamiltonian necessitates that the elements along the diagonal must be real, while the off-diagonal elements are complex. 

As we are only interested in coherent forward scattering of neutrinos in matter with zero momentum transfer, the SNSI parameters $\eta_{\alpha\beta}$ can be related to the parameters of the underlying theory as follows
\begin{align}\label{eq:mass_yukawa}
   \eta_{\alpha\beta} &= \dfrac{n_f y_f y_{\alpha\beta}}{\sqrt{|\Delta m_{31}^2|} m_\phi^2}   
\end{align}
Effectively, scalar NSIs induce a correction to $\mathbb{M}$ that enters $H_{ij} \propto \mathbb{M}^2_{ij}/2E_\nu$. Since the mapping between the fundamental Lagrangian parameters $y_{\alpha\beta}$ or the dimensionless $\eta_{\alpha\beta}$ to the impact on the propagation Hamiltonian (and therefore neutrino oscillations) is relatively complicated, we choose to introduce an effective parameterization for neutrino oscillations. We assume that $\mathbb{M}^2_{\rm eff}$ takes the following form
\begin{equation}\label{eq:msq_eff}
    \mathbb{M}_{\rm eff}^2 \equiv \left(\begin{array}{c c c} m_1^2 + \mu_{11} & \mu_{12} & \mu_{13} \\ \mu_{12}^* & m_2^2 + \mu_{22} & \mu_{23} \\ \mu_{13}^* & \mu_{23}^* & m_3^2 + \mu_{33} \end{array}\right).
\end{equation}
When $\mu_{ij} \to 0$, we recover the standard three-neutrino case. The diagonal $\mu$ are real and the off-diagonal are complex. We are insensitive to an overall constant times the identity appearing in this Hamiltonian, so we subtract $m_1^2 + \mu_{11}$ without loss of generality, recovering $\Delta m_{21}^2 + (\mu_{22} - \mu_{11})$ and $\Delta m_{31}^2 + (\mu_{33} - \mu_{11})$ along the diagonal.

We note here that oscillations are sensitive to eight additional real parameters when dealing with the most general scalar NSI scenario, i.e., not restricting ourselves to any specific flavor-structure in the $\phi$-$\nu$-$\nu$ interaction. However, nine real parameters appear in \cref{eq:deltaMalphabeta}, foreshadowing a redundancy when we interpret constraints on $\mu_{ij}$ with respect to $\eta_{\alpha\beta}$ -- we will address this in~\cref{sec:results}.\footnote{We also emphasize here that in standard three-flavor neutrino oscillations we are insensitive to the overall mass scale of the (ultrarelativistic) neutrinos $m_0$. However, when mapping between $\mu_{ij}$ and $\eta_{\alpha\beta}$ we will be, since $\mathbb{M}_{\rm eff}^2 = \left(\mathbb{M}_0 + \delta \mathbb{M}\right)^2$ will involve cross-terms between $\mathbb{M}_0$ and $\delta\mathbb{M}$.}

In the remainder of this work, we will study the capability of the upcoming DUNE experiment to test the scalar NSI paradigm by constraining the $\mu_{ij}$ parameterization. Subsequently, we will discuss how these constraints are related to flavor-specific constraints on $y_{\alpha\beta}$ and $m_\phi$ (from \cref{eq:mass_yukawa}) so that we can compare against (a) constraints pertaining directly to neutrino scattering and (b) constraints regarding light scalars coupling to SM charged fermions via the Yukawa couplings $y_f$.

\subsection{Mapping Between Parameter Spaces}\label{subsec:ParameterMapping}
In this section, we briefly discuss the parameter mapping back and forth between the $\eta_{\alpha\beta}$ basis -- directly related to the flavor-basis Yukawa couplings present in \cref{eq:mass_yukawa} -- adopted in the literature for scalar NSI and neutrino oscillations and our $\mu_{ij}$ basis in~\cref{eq:msq_eff}. First let us assume that $\delta \mathbb{M}_{ij} = U^\dagger \delta \mathbb{M}_{\alpha\beta} U$, i.e. we can rotate the flavor-basis modification to the neutrino masses in \cref{eq:deltaMalphabeta} into the mass basis by using the standard leptonic mixing matrix. 

We define $\delta\mathbb{M}_{ij}$ such that
\begin{equation}
 \delta \mathbb{M}_{ij} = U^\dagger\delta \mathbb{M}_{\alpha\beta} U =  \left(\begin{array}{lll} \epsilon_{11} & \epsilon_{12} & \epsilon_{13} \\ \epsilon_{12}^* & \epsilon_{22} & \epsilon_{23} \\ \epsilon_{13}^* & \epsilon_{23}^* & \epsilon_{33}
  \end{array}\right),
\end{equation}
so that $\epsilon_{ij}$ carry mass dimension (relative to the dimensionless $\eta_{\alpha\beta}$). We can then express $\mathbb{M}_{\rm eff}^2$ by taking $\left( \mathbb{M}_0 + \delta\mathbb{M}_{ij}\right)^2$, and setting elements equal. Comparing against~\cref{eq:msq_eff}, we obtain the following relationships
\begin{eqnarray} \label{eq:mapping}
\mu_{11}&=&2m_1\epsilon_{11}+\left|\epsilon_{11}\right|^2+\left|\epsilon_{12}\right|^2+\left|\epsilon_{13}\right|^2, \nonumber\\ 
    \mu_{22}&=& 2m_2\epsilon_{22}+\left|\epsilon_{22}\right|^2+\left|\epsilon_{12}\right|^2+\left|\epsilon_{23}\right|^2, \nonumber\\
    \mu_{33}&=& 2m_3\epsilon_{33}+\left|\epsilon_{33}\right|^2+\left|\epsilon_{13}\right|^2+\left|\epsilon_{23}\right|^2, \nonumber\\
    \mu_{12}&=& \left(m_1+m_2+\epsilon_{11}+\epsilon_{22}\right) \epsilon_{12}+\epsilon_{13} \epsilon_{23}^*,\nonumber\\
    \mu_{13}&=&\left(m_1+m_3+\epsilon_{11}+\epsilon_{33}\right) \epsilon_{13}+\epsilon_{12} \epsilon_{23}^*, \nonumber\\
    \mu_{23}&=&\left(m_2+m_3+\epsilon_{22}+\epsilon_{3 3}\right) \epsilon_{23}+\epsilon_{13} \epsilon_{12}^*.\label{eq:mapping}
\end{eqnarray}
For a given set of $\eta_{\alpha\beta}$, the $\epsilon_{ij}$ can be determined given a mixing matrix and a set of $m_1,$ $m_2$, $m_3$. Here we demonstrate the final mapping into $\mathbb{M}_{\rm eff}^2$ that enters the propagation Hamiltonian for neutrino oscillations. The inverse of this is less straightforward -- assuming an oscillation experiment can constrain the set of $\{\mu_{ij}\}$, interpreting those constraints instead as being on $\eta_{\alpha\beta}$ requires further assumptions. This is because oscillation experiments are only sensitive to the differences $\mu_{22} - \mu_{11}$ and $\mu_{33} - \mu_{11}$ instead of the overall scaling, so further assumptions on $\mu_{ij}$ (or equivalently $\epsilon_{ij}$) are required to solve the system of equations.

\subsection{Impact of SNSI on Long-Baseline Oscillation Probabilities}\label{subsec:Probabilities}
In the following sections, we will be interested in long-baseline experiments (such as DUNE) and their measurements of appearance ($\nu_\mu \to \nu_e$) and disappearance ($\nu_\mu \to \nu_\mu$) oscillation probabilities in a muon-neutrino beam. In this subsection, we inspect the impact of nonzero scalar NSIs on these oscillation probabilities, focusing on $P(\nu_\mu \to \nu_e)$ where the effects are most apparent.

\begin{table}[tbp]
\begin{center}
\begin{tabular}{|r||c|c|c|c|c|c|}\hline
& $\sin^2\theta_{12}$ & $\sin^2\theta_{13}$ & $\sin^2\theta_{23}$ & $\delta_{\rm CP}$ & $\Delta m_{21}^2$ & $\Delta m_{31}^2$ \\ \hline
Normal MO & $0.318$ & $0.022$ & $0.574$ & $1.08\pi$ & $7.506\times 10^{-5}$ eV$^2$ & $2.55 \times 10^{-3}$ eV$^2$ \\ \hline
Inverted MO & $0.304$ & $0.022$ & $0.578$ & $1.58\pi$ & $7.506\times 10^{-5}$ eV$^2$ & $-2.45\times 10^{-3}$ eV$^2$ \\ \hline
\end{tabular}
\caption{Oscillation parameters used as assumed-true values for our numerical studies~\cite{deSalas:2020pgw}.}
\label{table:oscpara}
\end{center}
\end{table}

In determining oscillation probabilities, we assume that the (constant) matter density of the earth along the path of propagation is $2.84$ g/cm$^3$ and that the baseline length $L = 1300$ km. Since the lightest neutrino mass factors in to the mapping between $\eta_{\alpha\beta}$ and $\mu_{ij}$ (\cref{eq:mapping}), we take the normal mass-ordering parameters in \cref{table:oscpara} and $m_1 = 0.1$ eV for demonstrative purposes.
 
\begin{figure}[tbp]
\centering 
\includegraphics[width=.48\textwidth]{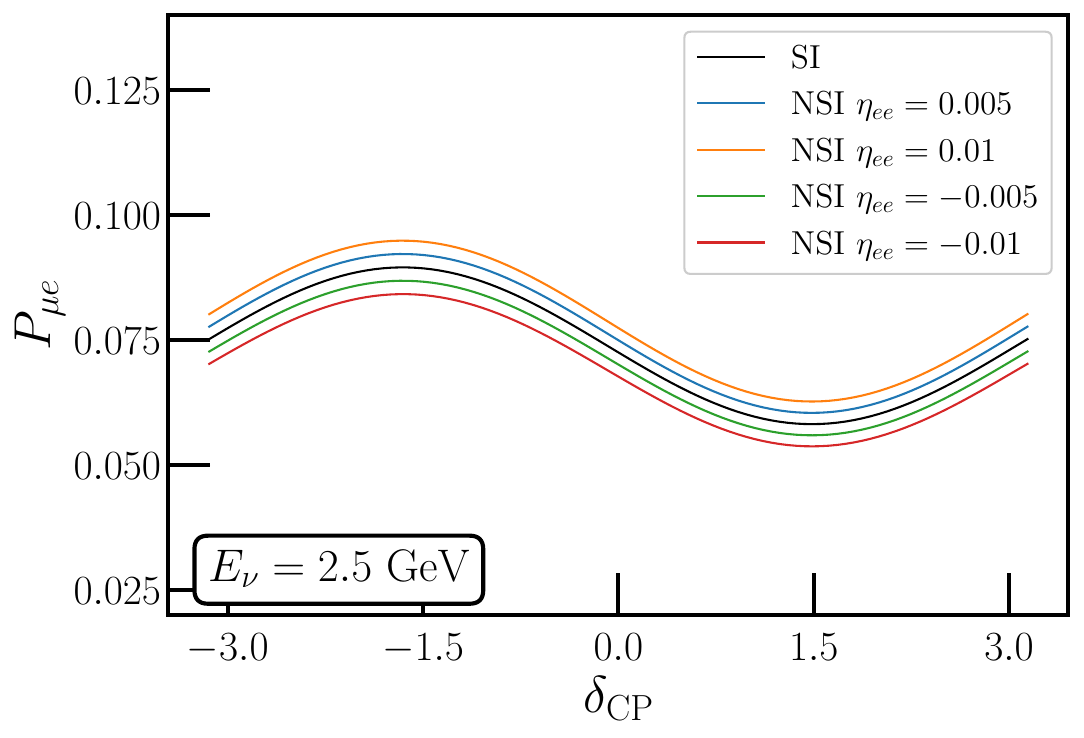}
\hfill
\includegraphics[width=.48\textwidth]{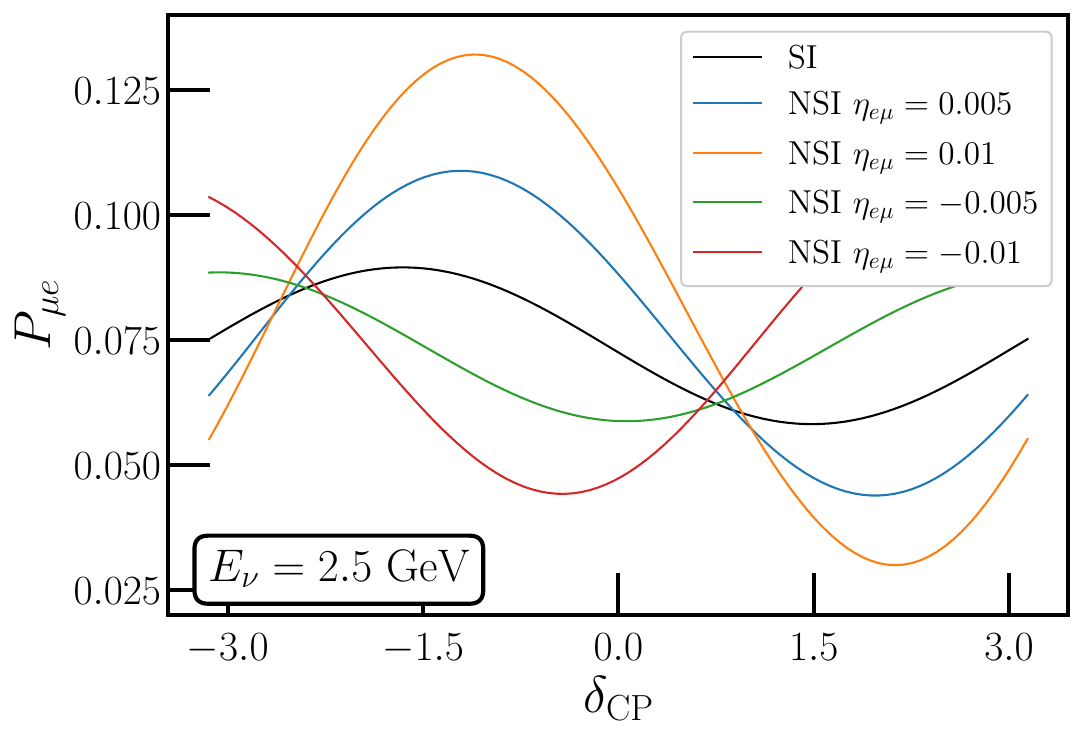}
\hfill
\caption{\label{fig:oscprobdcp} 
The effects of $\eta_{ee}$ (left) and $\eta_{e\mu}$ (right) on oscillation probabilities $P_{\mu e}$. We fix the normal mass ordering and parameters from~\cref{table:oscpara}, varying $\delta_{\rm CP}$ in each panel, and fix $E_{\nu} = 2.5$ GeV. The black line corresponds to the standard interaction case($\eta_{\alpha\beta}=0$) while other lines correspond to the presence of non-zero NSI parameters. Similar effects on oscillation probabilities exist for other $\eta_{\alpha\beta}$.}
\end{figure}

 In~\cref{fig:oscprobdcp}, we focus instead on the impact of scalar NSI and CP violation on the oscillation probability near its peak, $E = 2.5$ GeV. In the three-flavor scenario (even with matter effects), the oscillation probability can be shown to vary primarily with respect to $\sin\delta_{\rm CP}$ due to its dependence on the Jarlskog invariant. In contrast, we see now with $\eta_{\alpha\beta} \neq 0$, this trend can change (even with real NSI parameters). Such degeneracies between the scalar NSI parameters and $\delta_{\rm CP}$ will make robust extraction of the CP-violating parameters~\cite{Medhi:2021wxj,Medhi:2022qmu,Denton:2022pxt,Sarker:2023qzp,ESSnuSB:2023lbg}, as well as precision sensitivity to nonzero $\eta_{\alpha\beta}$, more challenging as we will see in the coming results.


\section{SNSI sensitivities at DUNE } \label{sec:DUNE}

The Deep Underground Neutrino Experiment (DUNE) is a forthcoming long-baseline neutrino experiment in the US, with ${\sim}$few GeV muon-flavor neutrinos produced via a $120$ GeV proton beam at Fermilab, travelling $1300$ km to South Dakota. The far-detector site in South Dakota will host large liquid-argon time-projection chamber detectors, with a target fiducial mass of 40 kiloton.\footnote{It is likely that this mass will not be achieved simultaneously with the onset of beam operation. We will quote our sensitivities in terms of kiloton$\times$MW$\times$year exposures, so that any rescaling for different detector mass/beam power/time may be performed.} We will perform analyses assuming a baseline beam power of $1.2$ MW and seven years of data collection, with equal operation in neutrino and antineutrino modes. This corresponds to an assumed exposure of $336$ kt$\times$MW$\times$yr. We refer the reader to Refs.~\cite{Berryman:2015nua,deGouvea:2015ndi,Berryman:2016szd,deGouvea:2016pom,deGouvea:2017yvn} for further details regarding how the event spectra in DUNE are simulated.

To evaluate DUNE's capability in constraining the scalar NSI parameters, we simulate the expected event rate (in four distinct oscillation channels) at DUNE assuming the standard oscillation parameters in~\cref{table:oscpara} and no new physics present. We then construct a test statistic $\chi^2 = -2\ln\mathcal{L}$ (where $\ln\mathcal{L}$ is the bin-by-bin Poissonian log-likelihood, summed over all experimental bins) by comparing this expected event rate against a rate determined using test parameters allowing the scalar NSI to be nonzero. We employ the Bayesian inference tool \texttt{MultiNest}~\cite{Feroz:2008xx,Buchner:2014nha} to study the parameter space and derive expected constraints on the new-physics parameters given expected DUNE data. In calculating the test statistic, we include 5\% normalization systematic uncertainties in each oscillation channel.

In~\cref{eq:msq_eff}, we see that there is also an unavoidable degeneracy between the ``vacuum'' $\Delta m_{j1}^2$ parameters and the diagonal $\mu_{jj}$ ones. Such degeneracies are present in different parameterizations of scalar NSI (e.g. the $\eta_{\alpha\beta}$ one) when considering multiple parameters simultaneously. To this end, we fix $\Delta m_{j1}^2$ to their best-fit values in our analyses and allow $(\mu_{jj} - \mu_{11})$ ($j = 2,$ $3$) to vary. The off-diagonal parameters and their CP-violating phases are also free to vary simultaneously, providing us with a thorough, conservative analysis of these future prospects at DUNE. The standard mixing angles $\sin^2\theta_{ij}$ and the CP-violating phase $\delta_{\rm CP}$ are also allowed to vary in our analyses. 

Using \texttt{MultiNest}, we obtain the expected constraints on all varied parameters -- in~\cref{app:CornerMu} we display the resulting corner plots demonstrating this sensitivity with respect to the $\mu_{ij}$ parameters assuming that the truth data are consistent with the normal mass ordering (\cref{fig:Cornerplt-NO-mu}) or the inverted mass ordering (\cref{fig:Cornerplt-IO-mu}). In~\cref{subsec:ParameterMapping}, we showed the nontrivial parameter mapping between $\mu_{ij}$ and $\eta_{\alpha\beta}$, including the dependence on the lightest neutrino mass $m_0$. To translate our results into constraints on $\eta_{\alpha\beta}$, we assume $m_0 = 0.1$ eV and $\epsilon_{11} = 0$, resulting in the corner plots shown in~\cref{fig:Cornerplt-NO} (NO) and~\cref{fig:Cornerplt-IO} (IO), also in~\cref{app:CornerMu}.  In~\cref{sec:results} we will discuss the impact of this $m_0$, $\epsilon_{11}$ assumption on the parameter mapping and on the results we obtain.

In comparing our results constraining $\mu_{ij}$ to those constraining $\eta_{\alpha\beta}$, it's worth noting the non-linearity of the conversion between the two parameterizations. This leads to an additional layer of degeneracy (because the $\eta_{\alpha\beta}$ modify $\mathbb{M}$ where the true sensitivity lies in $\mathbb{M}^2 \propto \mu_{ij}$) in determining constraints on $\eta_{\alpha\beta}$ when allowing all parameters to vary simultaneously. This leads to a degradation in DUNE sensitivity to scalar NSI. As a simple example, the transformation
\begin{eqnarray}
    \epsilon_{11}&\rightarrow& -\epsilon_{11}-2m_1 \nonumber \\
    \epsilon_{22}&\rightarrow& -\epsilon_{22}-2m_2 \nonumber \\ 
\epsilon_{33} &\rightarrow& -\epsilon_{33}-2m_3 \nonumber \\
\epsilon_{ij} &\rightarrow& -\epsilon_{ij};~~i \neq j
\end{eqnarray}leaves $\mathbb{M}^2$ and therefore the propagation Hamiltonian invariant. This leads to additional islands of degeneracy. In light of this, we attempt to prune these degenerate branches in order to consider the most optimistic constraints DUNE can hope to achieve, subject to this consideration. 


\section{Experimental bounds}
\label{Section:constraints}

In order to compare the DUNE projections against other experimental efforts, we must consider other searches attempting to discover new light scalars interacting with SM particles. For a light neutral scalar to impact terrestrial neutrino oscillations, it must couple to neutrinos as well as some earth matter, i.e. electrons or nucleons. This means that neutrino oscillations are sensitive to the combination $\sqrt{y_e y_\nu}$ or $\sqrt{y_N y_\nu}$ as a function of $m_\phi$, where other constraints may probe the individual Yukawa couplings as a function of $m_\phi$ independently.

For the remainder of this section, we will summarize the dominant constraints on these Yukawa interactions as a function of $m_\phi$. Then, in~\cref{sec:results} we will compare our DUNE sensitivity against these.

\paragraph{Fifth Force:} Scalar interactions between electrons or nucleons can lead to a modification from the expected Newton's law of gravitation on small scales (e.g.,~\cite{Damour:2010rm, Brzeminski:2022sde} etc.). Strong constraints on both $y_e$ and $y_N$, particularly for $m_\phi \lesssim$ 0.1 eV, have been derived from a lack of this deviation in various experiments. Specifically, torsion balance experiments~\cite{Schlamminger:2007ht} and space-based experiments for very small mass~\cite{Berge:2017ovy, MICROSCOPE:2022doy} provide the strongest constraints.

 \paragraph{CMB Observations:} Neutrino-scalar interactions can have multiple effects in the early universe, including delaying neutrino free-streaming and adding additional, thermalized, light degrees of freedom around the epoch of the formation of the Cosmic Microwave Background. Constraints of $\Delta N_{\rm eff}$ at this epoch lead to very strong constraints on $y_\nu$ down to very small $m_\phi \lesssim$ eV~\cite{Sandner:2023ptm, NeffEscudero}.

\paragraph{BBN:} Similar to the above, such new mediators can spoil the successful prediction of light element abundances formed during big-bang nucleosynthesis. This can be spoiled via a variety of Yukawa couplings, either among electrons (modifying annihilation processes during this epoch) or neutrinos (modifying their free-streaming as above). We refer the reader to Ref.~\cite{Babu:2019iml} for further discussion of these effects, and use the constraints therein.

\paragraph{Stellar and Supernova Cooling:} New, light mediators can be emitted in stellar environments, leading to significantly higher cooling rates than expected. The strongest constraints of this type come from red giant and horizontal branch stars~\cite{Hardy:2016kme} for very light mediators, requiring $y_e \lesssim 10^{-15}$ and $y_N \lesssim 10^{-12}$~\cite{Babu:2019iml}. Supernovae provide constraints from similar arguments, and the observation of neutrinos from SN1987A allow one to exclude the regions $10^{-9} \lesssim y_e\lesssim 10^{-7}$ and $10^{-10} \lesssim y_N\lesssim 10^{-7}$ if $m_\phi < m_e$~\cite{Babu:2019iml}.

\paragraph{In-medium Neutrino Mass:} Just as the scalar NSI contribute to an effective mass term for neutrino oscillations, they can generate an effective mass for neutrinos produced or propagating in a sufficiently dense environment. Constraints on the effective mass of neutrinos produced in dense environments (particularly the sun and in a supernova explosion) have been translated into constraints on the Yukawa couplings and mediator mass in Ref.~\cite{Babu:2019iml}, and yield some of the most severe bounds for very small mediator masses. Specifically, Borexino data~\cite{Borexino:2017rsf} were used to put bounds in the case of sun, while data from the SN1987A event~\cite{Smirnov:2019cae, Babu:2019iml} were used to set constraints in the case of supernova. We will discuss the possibility of relaxing such constraints in~\cref{sec:results} when placing our results in context.

\paragraph{Neutrino Scattering Experiments:} Neutrino scattering off nuclei (and electrons nearby) allow experiments to constrain the product $y_\nu y_{e/N}$. Various lab based experiments: elastic neutrino-electron scattering such as TEXONO~\cite{TEXONO:2009knm}, BOREXINO~\cite{BOREXINO:2018ohr}, GEMMA~\cite{Beda:2013mta}, XENONnT~\cite{XENON:2022ltv}, LZ~\cite{LZ:2022lsv} and CEvNS experiment such as COHERENT~\cite{Akimov:2017ade, Akimov:2018vzs, Akimov:2018ghi, Akimov:2019xdj, Akimov:2020pdx, COHERENT:2021pvd} put bounds on this through measurements of neutrino-electron/neutrino-nucleon elastic scattering rates. In neutrino-electron scattering experiments, XENONnT provides the most stringent constraint, restricting $\sqrt{y_\nu y_{e}} \lesssim8 \times 10^{-7}$ for $m_\phi \lesssim 10$ keV. In the case of neutrino-nucleus scattering, COHERENT has placed bounds on $\sqrt{y_\nu y_{N}} \lesssim 2 \times 10^{-5}$ for $m_\phi \lesssim 10$ MeV.

Note that only the neutrino scattering experiments can put constraints on the product coupling $\sqrt{y_\nu y_{e/N}}$, while all other constraints are on the individual couplings. To obtain constraints on the product coupling $\sqrt{y_\nu y_{e/N}}$, we combine the individual constraints on $y_{e/N}$ and $y_\nu$.
 
\section{Results} \label{sec:results}
As we begin assessing the results of our DUNE scalar NSI analysis, we first address the question of CP violation when considering an analysis with more parameters than in the standard oscillation case. In~\cref{fig:oscprobdcp} we demonstrated the degeneracy between allowing nonzero SNSI parameters and the standard CP violating phase $\delta_{\rm CP}$, alluding to the fact that analyses of this sort will lead to diminished power of determining this parameter -- Ref.~\cite{Medhi:2022qmu} explored some aspects of this question.

\begin{figure}[tbp]
\centering 
\includegraphics[width=0.65\textwidth]{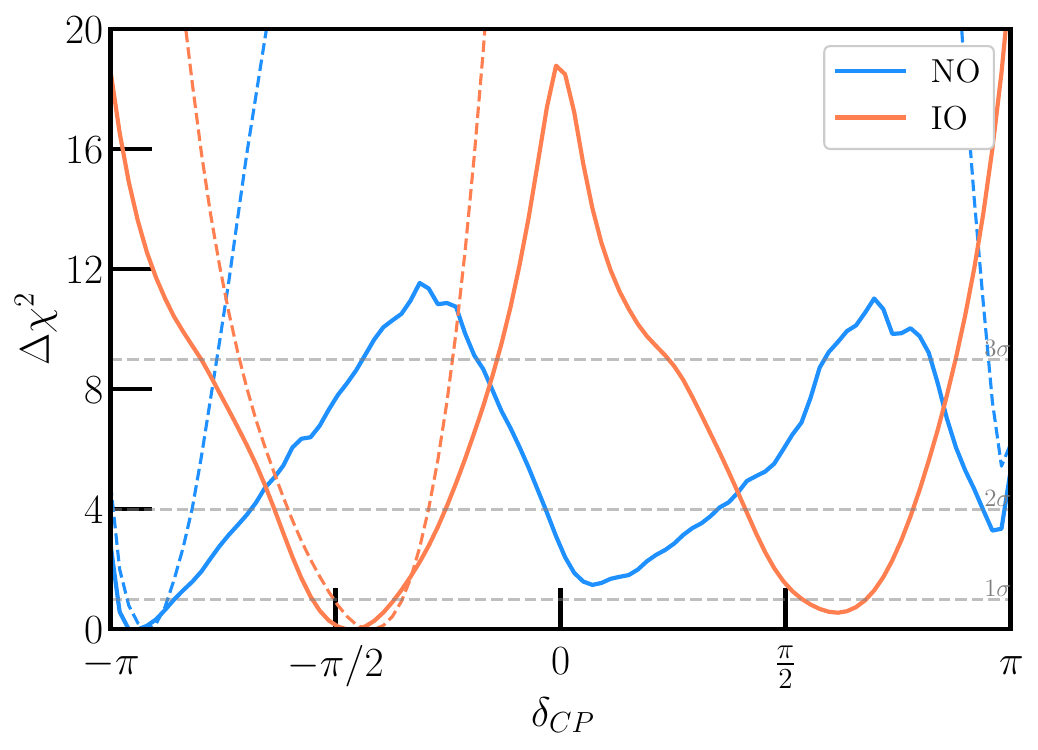}
\hfill
\caption{\label{fig:deltachisqvsdcp} 
DUNE's potential towards CP precision measurements in the presence of SNSI considering all parameters at the same time taking $m_{\text{lightest}}=0.1$ eV, $\epsilon_{11}=0.0 ( \epsilon_{33}=0.0)$ for NO (IO). The solid blue (orange) line corresponds to the NO (IO) case in the presence of SNSI while the dashed blue (orange) line is the corresponding curve in the absence of any SNSI.  }
\end{figure}
\cref{fig:deltachisqvsdcp} demonstrates the reduction in measurement power of $\delta_{\rm CP}$ assuming either set of true parameters given in~\cref{table:oscpara}. All other parameters in our analysis (see~\cref{sec:DUNE} for details) are marginalized in constructing this figure. Here, the blue (orange) curves represent our expected measurement potential by DUNE assuming that nature has chosen the normal (inverted) mass ordering. The solid lines represent our analysis considering all SNSI parameters to be simultaneously nonzero, whereas the dashed ones are for a ``three-flavor-only'' analysis, i.e. $\mu_{ij} = 0$.
For each truth scenario, we see a significant reduction in the ability to measure $\delta_{\rm CP}$, as well as the capability to exclude $\delta_{\rm CP} = 0$, $\pi$, i.e. to determine whether or not CP is conserved in the lepton sector. We highlight here too that additional local minima appear ($\delta_{\rm CP}\sim0$ for NO and $\delta_{\rm CP}\sim\pi/2$ for IO) in these measurement plots due to the degeneracy between ``fake'' solutions when nonzero SNSI are allowed. We note however, one advantage of having several next-generation experiments that are uniquely sensitive to SNSI. Combined analyses, for instance with JUNO~\cite{Gupta:2023wct} and Hyper-Kamiokande, should in principle guard against such dramatic parameter degeneracies and allow for robust measurements/constraints on all parameters. We leave a dedicated study of such combinations to future work.

\begin{figure}[!htbp]
    \centering
    \includegraphics[width=0.48\textwidth]{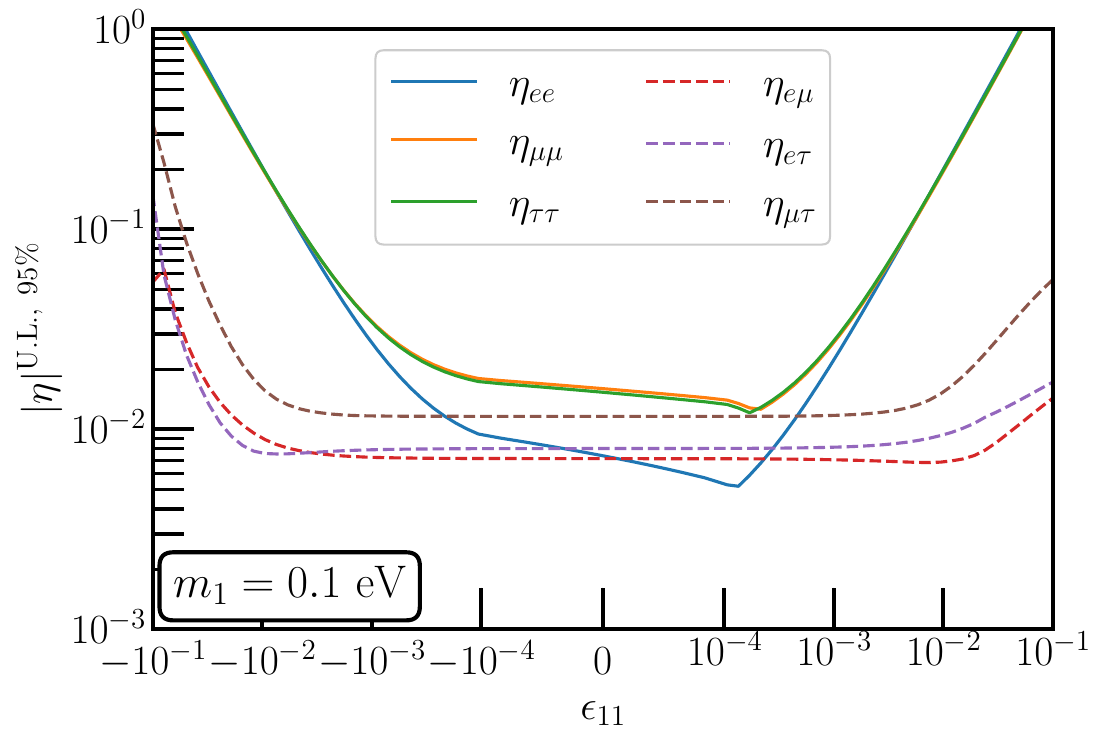}
    \includegraphics[width=0.48\textwidth]{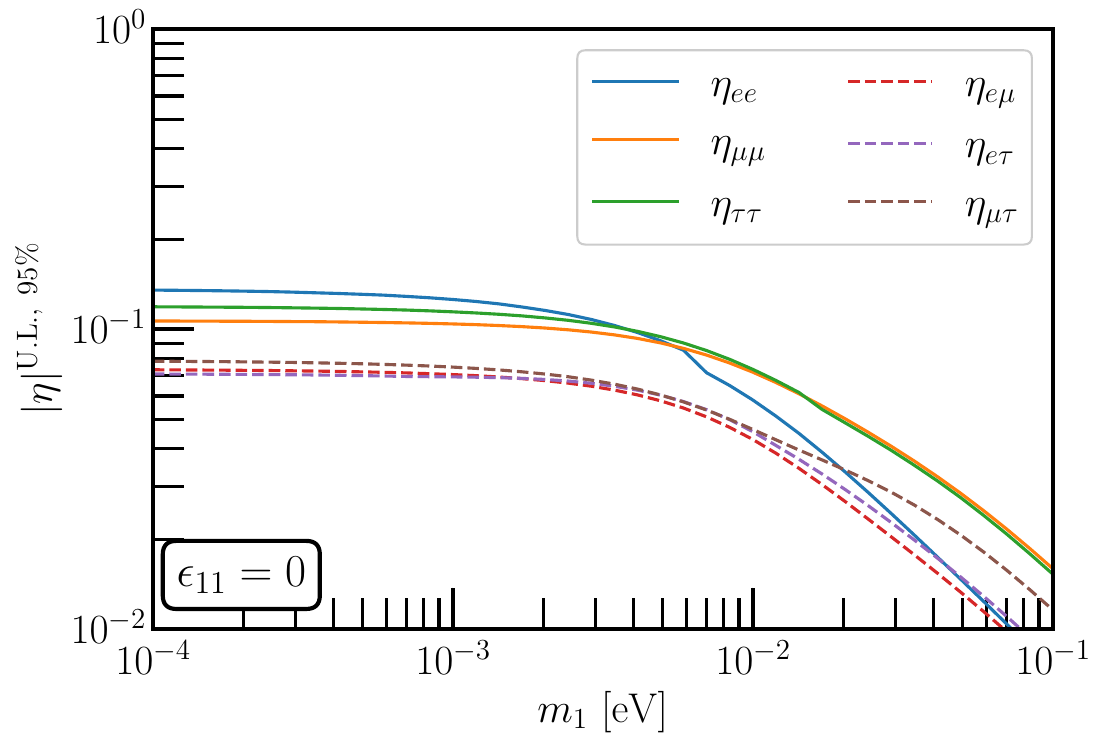}
    \caption{Left: Upper limits  of various SNSI parameters $\eta_{\alpha\beta}$ as a function of $\epsilon_{11}$ at 95\% C.L. for the normal mass ordering case with the lightest mass taken as $m_1=0.1$~eV. Similar trends would follow for the inverted mass ordering scenario in relation to the variable $\epsilon_{33}$ for $m_3 = 0.1 $ ~eV.
    Right: Same but fixing $\epsilon_{11} = 0$ and varying $m_1$.}
    \label{fig:eta_vs_eps11_m1_NO} 
\end{figure}

Next, it is our goal to take the results of our constraints on the $\mu_{ij}$ parameters (see~\cref{app:CornerMu} for corner plots and more discussion) into constraints on the oft-studied $\eta_{\alpha\beta}$ ones. First, as a benchmark for comparison, we perform analyses studying the sensitivity to nonzero $\eta_{\alpha\beta}$ one-parameter-at-a-time. Such analyses require a choice of $m_1$ (or some overall mass scale), and we set $m_1 = 0.1$ eV ($m_3 = 0.1$ eV) when considering the NO (IO). We provide the expected sensitivity at 95$\%$ Confidence Level (C.L.) to the various $\eta_{\alpha\beta}$ in this approach in~\cref{table:expsensone}.
\begin{table}[!htbp]
\centering
\begin{tabular}{|c||c|c|}\hline
NSI & NO ($m_1$ = 0.1 eV) & IO ($m_3 = 0.1$ eV) \\ \hline \hline
$\eta_{ee}$ & $[-0.006,0.006]$ & $[-0.0077,0.0059]$ \\ \hline
$\eta_{\mu\mu}$ & $[-0.004,0.004]$ & $[-0.0036,0.004]$ \\ \hline
$\eta_{\tau\tau}$ & $[-0.004,0.004]$ & $[-0.0037,0.0036]$ \\ \hline
$|\eta_{e\mu}|$ & [0,0.0017] & [0,0.0017] \\ \hline
$|\eta_{e\tau}|$ & [0,0.0019] & [0,0.0015] \\ \hline
$|\eta_{\mu\tau}|$ & [0,0.0042] & [0,0.0035] \\ \hline
\end{tabular}
\caption{Summary of expected sensitivities of DUNE to SNSI parameters $\eta_{\alpha\beta}$ at 95\% C.L. when considered one at a time for $m_{\text{lightest}} = 0.1$~ eV. }
\label{table:expsensone}
\end{table}

When allowing all parameters to vary simultaneously (as discussed in~\cref{sec:scalarNSI}), the larger number of free parameters in the $\eta_{\alpha\beta}$ parameterization relative to the $\mu_{ij}$ one means that choices must be made in this mapping. In addition to fixing the lightest neutrino mass, this amounts to choosing $\epsilon_{11}$ ($\epsilon_{33}$) in the NO (IO), one of the mass-basis SNSI parameters. In~\cref{fig:eta_vs_eps11_m1_NO}(left), we demonstrate how the constraints on $\mu_{ij}$ from our normal-mass-ordering analysis translate into those on $\eta_{\alpha\beta}$ at $95\%$ CL, when $\epsilon_{11}$ is varied and $m_1$ is fixed to be $0.1$ eV. The results when assuming the inverted mass ordering are qualitatively similar. In both cases, when $\epsilon_{11}$ (or $\epsilon_{33}$ is allowed to be large, degeneracies in the mapping between parameter spaces (see~\cref{eq:mapping}) allow several parameters to be simultaneously large leading to large $\eta_{\alpha\beta}$ (note that here we are presenting upper-limit constraints on $|\eta_{\alpha\beta}|$, where the degeneracies may rely on the diagonal $\eta_{\alpha\alpha}$ being negative and/or the off-diagonal ones being complex). 

We can repeat the same procedure, fixing $\epsilon_{11}$ (or $\epsilon_{33}$ in the IO) and varying the lightest neutrino mass $m_1$ ($m_3$), and analyze how the constraints on $\eta_{\alpha\beta}$ vary. This is presented in~\cref{fig:eta_vs_eps11_m1_NO} (right) for the six different magnitudes under the NO assumption and $\epsilon_{11} = 0$. Again, the qualitative results are similar when considering the IO and varying $m_3$ -- we find that the most optimistic constraints we can hope to obtain by DUNE are when $\epsilon_{11}$ ($\epsilon_{33}$) is zero and $m_1$ ($m_3$) is taken to be as large as reasonably possible, approximately $0.1$ eV. However, the constraints, unless we end up in a highly-degenerate region, do not get more than an order-of-magnitude or so worse. For these reasons, we will take the optimistic scenario that the lightest neutrino mass is $0.1$ eV and $\epsilon_{11} = 0$ ($\epsilon_{33} = 0$) in the NO (IO).

In comparison to~\cref{table:expsensone}, we provide expected constraints on the $\eta_{\alpha\beta}$ when \textit{all} parameters are allowed to vary simultaneously in~\cref{table:expsens}. We do so for three demonstrative choices for the lightest neutrino mass, $0$ eV, $0.05$ eV, and $0.1$ eV (where the expected constraints tend to improve with increasing $m_{\rm lightest}$, as demonstrated in~\cref{fig:eta_vs_eps11_m1_NO} right).
\begin{table}[!htbp]
\begin{minipage}[c]{0.33\textwidth}
\centering
$$
\begin{array}{|c|c||c|}
\multicolumn{3}{c}{m_{\text {lightest }}=0 ~\text{eV}} \\
\hline
\text { MO } & \text { NSI } & \eta_{\alpha \beta}  \\
\hline
&\eta_{ee}                  & [-0.133,0.063]  \\

&\eta_{\mu\mu}              & [-0.107,0.059] \\

\text{NO}&\eta_{\tau\tau}            & [-0.12,0.075]  \\

&\eta_{e\mu}                & [0,0.074]   \\

&\eta_{e\tau}               & [0,0.072]   \\

&\eta_{\mu\tau}             & [0,0.079]  \\
\hline
& \eta_{ee} & [-0.017,0.019] \\
& \eta_{\mu \mu} & [-0.028,0.044] \\
\text { IO } & \eta_{\tau \tau} & [-0.031,0.041] \\
& \eta_{e \mu} & [0,0.025] \\
& \eta_{e \tau} & [0,0.024] \\
& \eta_{\mu \tau} & [0,0.052] \\
\hline
\end{array}
$$
\end{minipage}
\begin{minipage}[c]{0.33\textwidth}
\centering
$$
\begin{array}{|c|c||c|}
\multicolumn{3}{c}{m_{\text {lightest }}=0.05 ~\text{eV}} \\
\hline
\text { MO } & \text { NSI } & \eta_{\alpha \beta}  \\
\hline
& \eta_{ee} & [-0.015,0.005] \\
& \eta_{\mu \mu} & [-0.028,0.015] \\
\text { NO } & \eta_{\tau \tau} & [-0.027,0.016] \\
& \eta_{e \mu} & [0,0.013]\\
& \eta_{e \tau} & [0,0.015] \\
& \eta_{\mu \tau} & [0,0.02] \\
\hline
& \eta_{ee} & [-0.012,0.013] \\
& \eta_{\mu \mu} & [-0.019,0.012] \\
\text { IO } & \eta_{\tau \tau} & [-0.014,0.017] \\
& \eta_{e \mu} & [0,0.012] \\
& \eta_{e \tau} & [0,0.013] \\
& \eta_{\mu \tau} & [0,0.021] \\

\hline
\end{array}
$$
\end{minipage}
\begin{minipage}[c]{0.33\textwidth}
\centering
$$
\begin{array}{|c|c||c|}
\multicolumn{3}{c}{m_{\text {lightest }}=0.1~ \text{eV}} \\
\hline
\text { MO } & \text { NSI } & \eta_{\alpha \beta}  \\
\hline
& \eta_{ee} & [-0.0074,0.0026]  \\
& \eta_{\mu \mu} & [-0.016,0.009] \\
\text { NO } & \eta_{\tau \tau} & [-0.016,0.008] \\
& \eta_{e \mu} & [0,0.007] \\
& \eta_{e \tau} & [0,0.008] \\
& \eta_{\mu \tau} &[0,0.016]  \\
\hline
& \eta_{ee} & [-0.0077,0.0082] \\
& \eta_{\mu \mu} & [-0.011,0.0067] \\
\text { IO } & \eta_{\tau \tau} & [-0.0081,0.0096] \\
& \eta_{e \mu} & [0,0.0073] \\
& \eta_{e \tau} & [0,0.0077] \\
& \eta_{\mu \tau} & [0,0.012]\\
\hline
\end{array}
$$
\end{minipage}
\caption{Summary of expected sensitivities of DUNE to SNSI parameters $\eta_{\alpha\beta}$ at 95\% C.L. when considered all at the same time for $\epsilon_{11}=0.0 $(NO) and  $\epsilon_{33}=0.0 $(IO)}
\label{table:expsens}
\end{table}

\begin{figure}[t]	
\centering
\begin{subfigure}{0.48\textwidth}\includegraphics[width=0.97\linewidth]{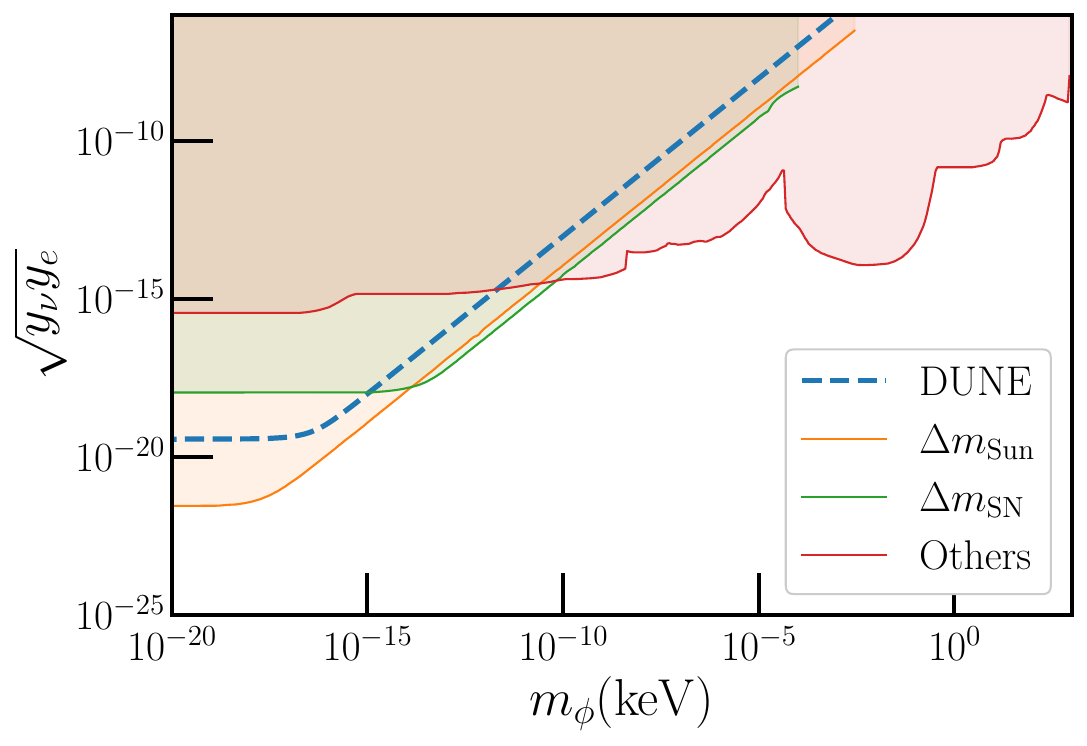}
\caption{\label{fig:constpltnue1}}							\end{subfigure}	
\begin{subfigure}{0.48\textwidth}		\includegraphics[width=0.97\linewidth]{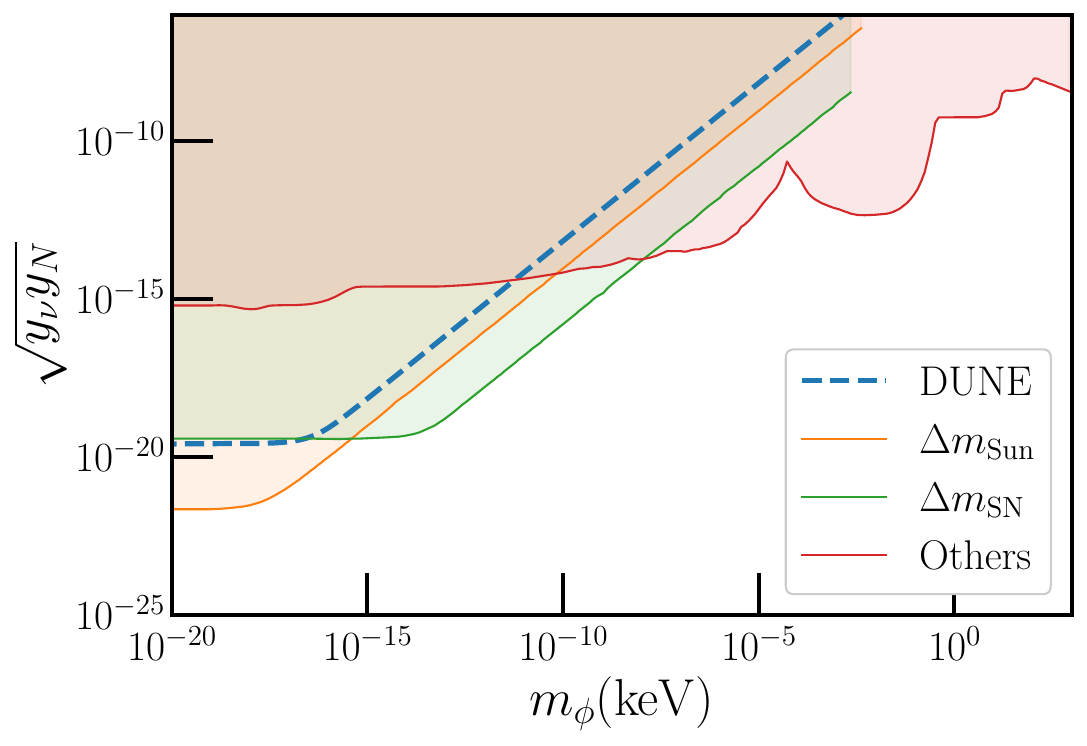}				
\caption{\label{fig:constpltnuq1}}							\end{subfigure}	
	
\captionsetup{justification   = RaggedRight,
             labelfont = bf}
\caption{\label{fig:constpltnue} The sensitivity of DUNE experiment at 95\% C.L. (dashed blue) on the light scalar mediator $\phi$ mass and couplings in the $(m_\phi,\sqrt{y_\nu y_{e/N}})$ plane. The left panel is for electron couplings $\sqrt{y_\nu y_e}$ while the right panel is for nucleon couplings $\sqrt{y_\nu y_N}$. We assume the NO, $m_1 = 0.1$ eV, and $\epsilon_{11} = 0$ (constraints are nearly identical in the IO). Also shown are other relevant constraints as shaded region, which are based on a single NSI parameter while the DUNE sensitivity incorporates the full set of NSI parameters. In particular the ``Others'' represents combination of constraints on $y_{e/N}$ and $y_\nu$ coming from BBN, $\Delta N_{\rm eff}$, fifth force search, stellar and supernova cooling. $\Delta m_{\rm Sun}$ ($\Delta m_{\rm SN}$) represents constraints due to the in-medium mass of the neutrinos inside the Sun (SN1987A).}
\end{figure}

Our final goal is to evaluate how DUNE's sensitivity to SNSI parameters compares with that of various non-oscillation-based probes mentioned in section.~\ref{Section:constraints}. To facilitate a meaningful comparison, we begin by converting DUNE's projected sensitivities on SNSI parameters into projected sensitivities on the model parameters, which include Yukawa couplings and mass. This translation is achieved using the relationship established in ~\cref{eq:mass_yukawa}. Specifically, we consider  values for the $\eta_{\alpha\beta}$ parameters that correspond to the $95\%$ C.L. contours depicted in Fig.~\ref{fig:Cornerplt-NO} and Fig.~\ref{fig:Cornerplt-IO} for the NO and IO scenarios, respectively. Note that these contours correspond to the scenario where the lightest neutrino mass is fixed at $0.1$~eV. We then superimpose these projected lines onto the bounds obtained from various terrestrial experiments, as well as astrophysical and cosmological probes in coupling vs mass plane $(\sqrt{y_f y_{\alpha \beta}}, m_\phi)$. Specifically, in Figure ~\ref{fig:constpltnue1}, we present the results for electron couplings, while in Figure ~\ref{fig:constpltnuq1}, we show the outcomes for nucleon couplings.

This combined representation provides a comprehensive view of the parameter space. We observe that the parameter space that DUNE will robustly explore with its oscillation programme appears to already be constrained by current data from various non-oscillation probes. We note that, this being the first study that allows for all SNSI parameters varying simultaneously, cancellations have occurred that limit DUNE's sensitivity in our approach. This can be seen by comparing the relative sensitivities presented in~\cref{table:expsensone} (one-parameter-at-a-time) with~\cref{table:expsens} (all-varying-simultaneously). If this high-dimension parameter space is explored thoroughly in the context of the other competing constraints, a decent relaxation of those is expected as well -- especially when considering the in-medium contribution to effective masses shown as $\Delta m_{\rm Sun}$ and $\Delta m_{\rm SN}$ in~\cref{fig:constpltnue}. With this in mind, we find it likely that DUNE can contribute new information regarding SNSI with very light mediator masses, e.g. $m_\phi \lesssim 10^{-12}$ keV.

Recent tensions in the T2K and NOvA data in the measurement of the CP-violating phase $\delta_{CP}$ have prompted a reevaluation of our understanding of neutrino oscillations. One possible solution to reconcile these tensions is to consider the effects of complex non-diagonal NSIs as has been in shown in Ref.~\cite{Denton:2022pxt}. However, our analysis suggests that the parameter space required to explain these tensions using SNSIs is largely ruled out by the existing constraints. The vector case, on the other hand, could potentially accommodate the required parameter space that explains this discrepancy. In our research, we are primarily interested in renormalizable gauge theory models. Consequently, we cannot construct a vector boson model with CP-violating complex coupling. So this study is not applicable for vector-mediated NSI. However, it's worth noting that theories with an effective Lagrangian can accommodate such complex couplings which provide a framework to study the CP-violating effects in the context of vector NSIs.


\section{Conclusion} \label{sec:conclusion}

 We have studied NSIs between neutrinos and SM fermions mediated by light neutral scalar fields. Neutrino models incorporating light scalar fields can accommodate such new interactions.  The parameter space associated with these light scalars is constrained by various non-oscillation probes, including neutrino-electron (nucleon) scattering, fifth force searches, CMB studies, BBN, supernova and stellar cooling observations. Moreover, the presence of these new interactions would also influence the in-medium masses of neutrinos within dense environments like supernovae and the Sun.

 Another natural method for probing these new interactions is through neutrino oscillation experiments. The presence of scalar NSIs emerges as a correction to the neutrino mass matrix in the neutrino propagation Hamiltonian. The oscillation phenomenology of scalar NSIs exhibit distinct characteristics compared to the SM weak interactions of neutrinos and vector boson-mediated NSIs. As the effect of scalar NSI scales linearly with the matter density of the medium, long-baseline oscillation experiments are suitable candidates for their exploration. In our analysis, we  utilized DUNE as a case study.  In particular, scalar NSIs can mimic the CP effect and hinder the measurement efforts of the leptonic Dirac CP phase at DUNE.

 We introduced a new model-independent parameterization scheme to study the effect of SNSI that enables us to consider more than one nonzero SNSI parameters at a time. We performed a thorough analysis to find DUNE sensitivity on the SNSI parameter space when all the parameters were nonzero. We transformed these sensitivities on the SNSI parameters into projected sensitivities in the parameter space of the underlying light scalar models and compared them with the existing bounds. We found that the scalar parameter space, which DUNE can probe, is entirely constrained by current data. A comprehensive analysis considering all the SNSI parameters simultaneously could potentially unveil unexplored parameter space in the low mass regime, for instance,  $m_\phi \lesssim 10^{-12}$~keV or so. 

 As scalar NSIs depend on the matter density of the medium, a synergy between various long baseline experiments would be highly beneficial in enhancing our understanding of the parameter space associated with SNSI and guarantee a correct measurements of CP phase angle. Moreover, a reevaluation of the existing constraints within a higher dimensional parameter space, considering all the SNSI parameters simultaneously, would facilitate comparisons between DUNE's sensitivity and other non-oscillatory probes.

\acknowledgments

  B.D., A.T., K.J.K. and A.V. receive partial support from DOE grant DE-SC0010813. S.G.'s work is funded by the National Research Foundation of Korea (NRF) Grant No. NRF-2019R1A2C3005009 (SG). T.L. acknowledges support from the National Key Research and Development Program of China Grant No. 2020YFC2201504, by the Projects No. 11875062, No. 11947302,  No. 12047503, and No. 12275333 supported by the National Natural Science Foundation of China, as well as  the Key Research Program of the Chinese Academy of Sciences, Grant NO. XDPB15. We also recognize that portions of this research were conducted using the advanced computing resources provided by Texas A\&M High Performance Research Computing.

\appendix

\

\section{\texttt{MultiNest} Corner plots}\label{app:CornerMu}

\begin{figure}[tbp]
\centering 
\includegraphics[width=\textwidth]{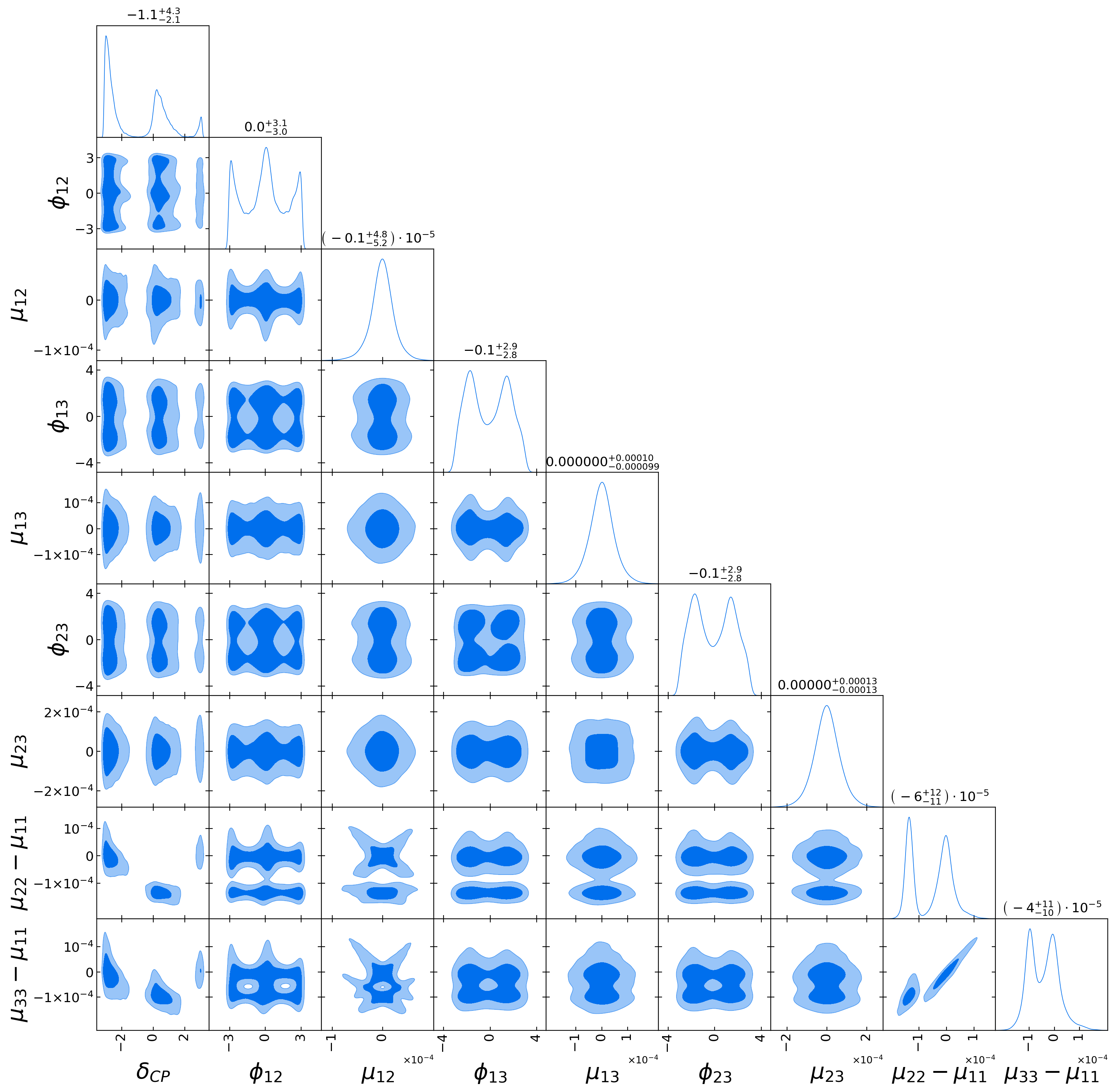}
\hfill
\caption{\label{fig:Cornerplt-NO-mu}  Corner plot showcasing DUNE sensitivity to the $\mu_{ij}$ parameters where the truth data are simulated assuming no new physics and according to the normal mass ordering.}
\end{figure}

\begin{figure}[tbp]
\centering 
\includegraphics[width=\textwidth]{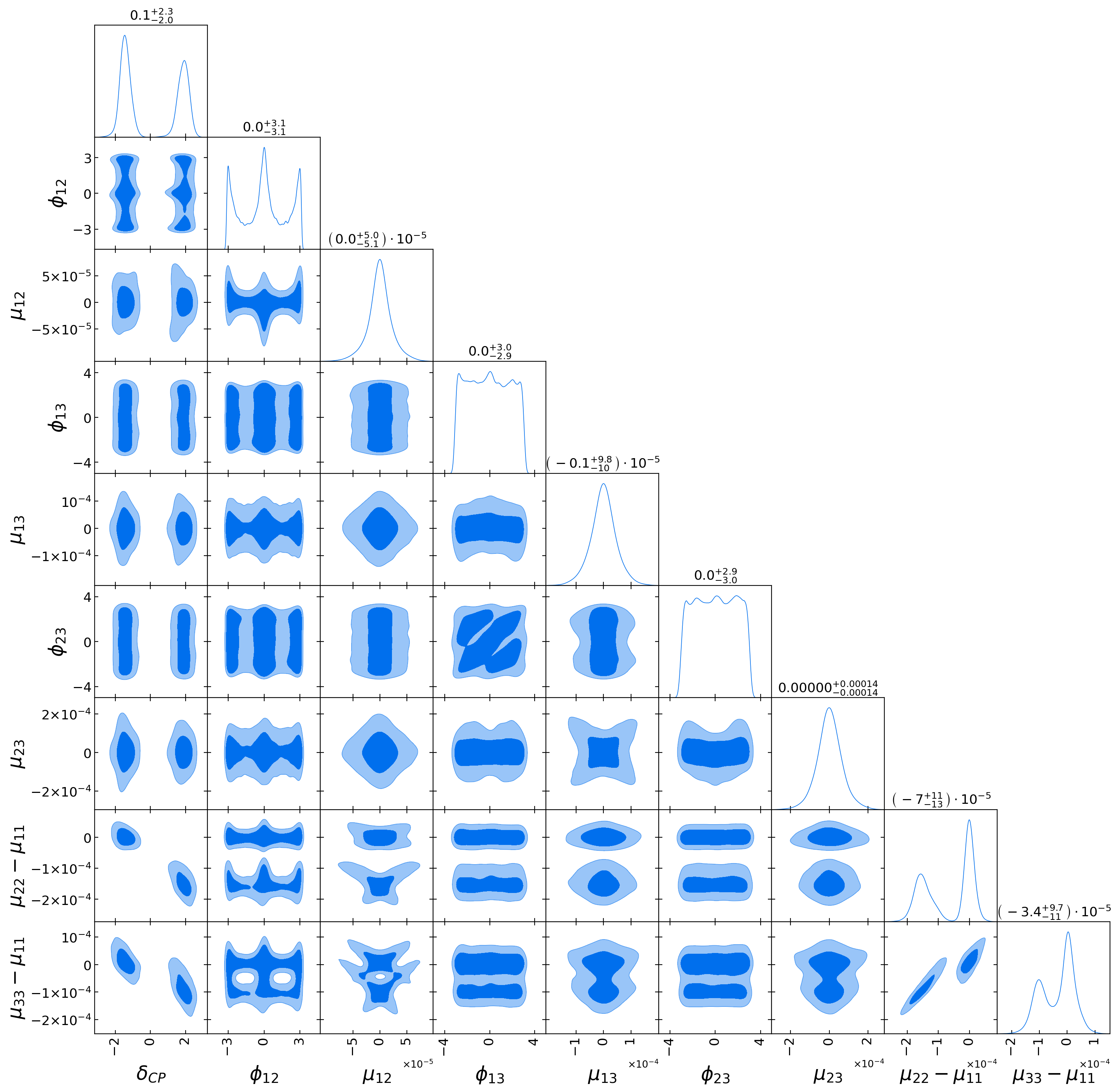}
\hfill
\caption{\label{fig:Cornerplt-IO-mu} Corner plot showcasing DUNE sensitivity to the $\mu_{ij}$ parameters where the truth data are simulated assuming no new physics and according to the inverted mass ordering.}
\end{figure}

\begin{figure}[tbp]
\centering 
\includegraphics[width=\textwidth]{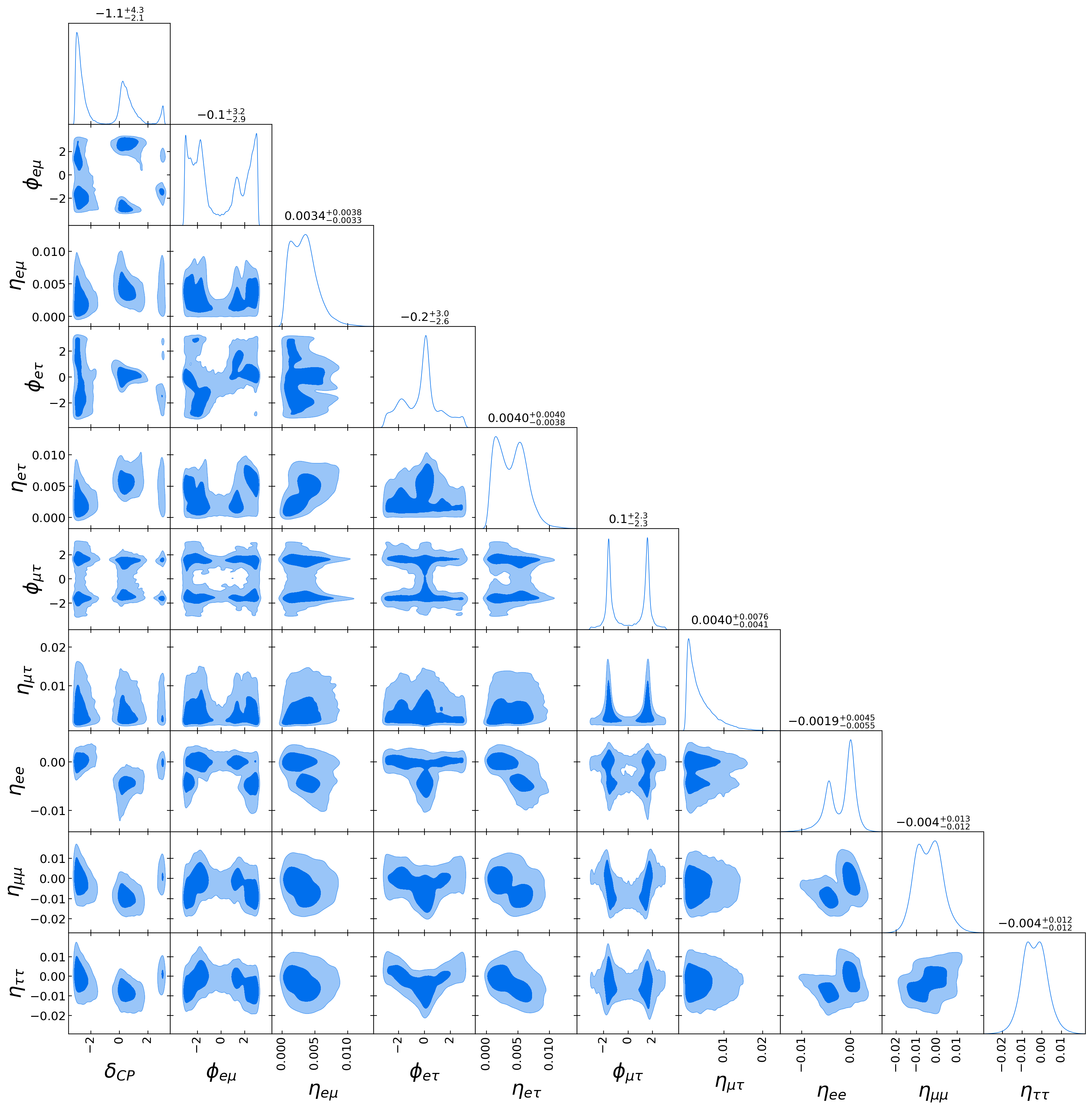}
\hfill
\caption{\label{fig:Cornerplt-NO} Corner plot demonstrating the sensitivity to the $\eta_{\alpha \beta}$ parameters for normal mass ordering. While obtaining these sensitivities, we have set the following parameters as fixed: $m_{1}=0.1$ eV and $\epsilon_{11}=0$ eV.}
\end{figure}

\begin{figure}[tbp]
\centering 
\includegraphics[width=\textwidth]{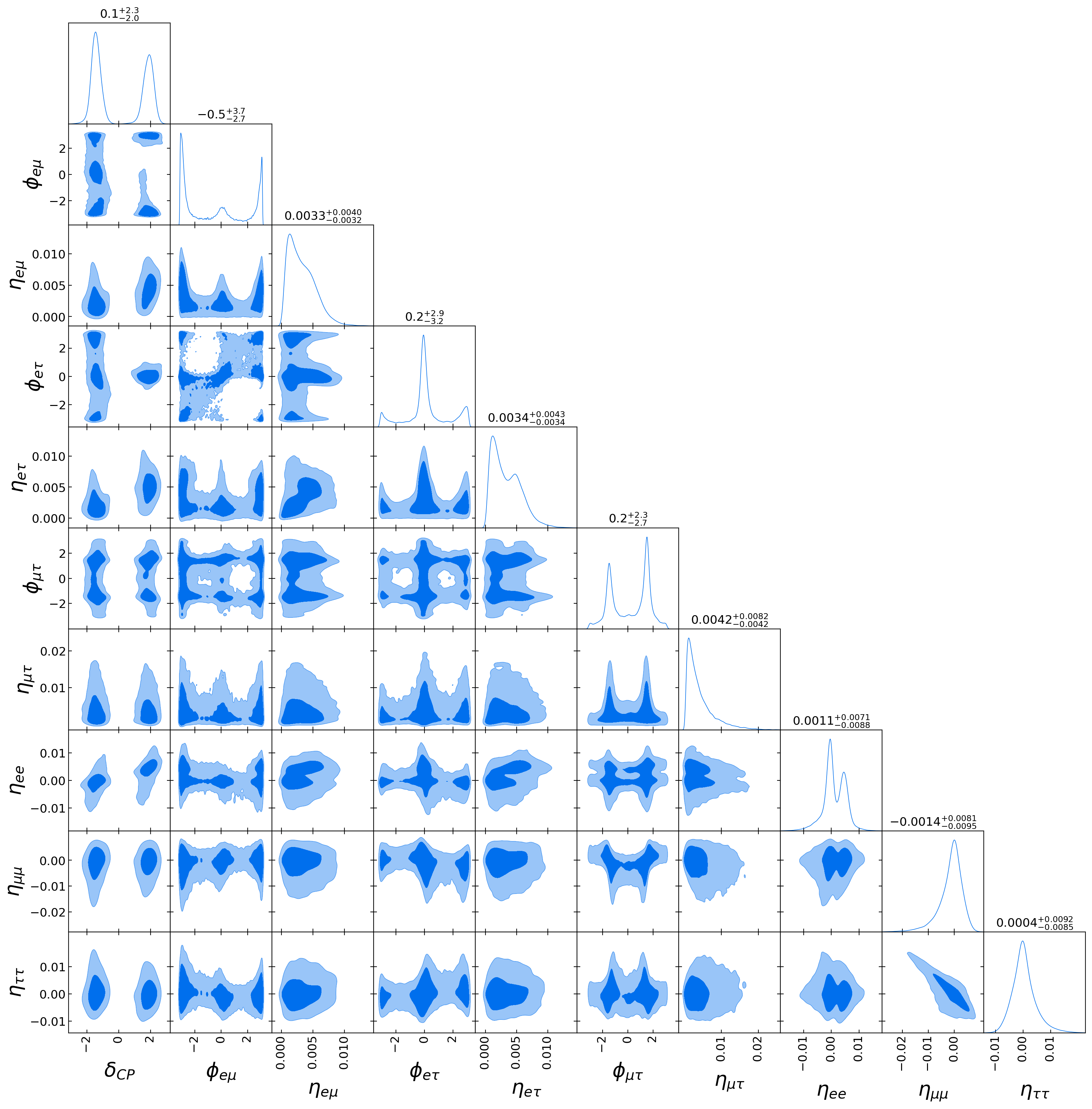}
\hfill
\caption{\label{fig:Cornerplt-IO} Corner plot demonstrating the sensitivity to the $\eta_{\alpha \beta}$ parameters for inverted mass ordering. While obtaining these sensitivities, we have set the following parameters as fixed:
$m_3=0.1$  eV and $\epsilon_{33}=0$ eV.}
\end{figure}

Here we show the results of our \texttt{MultiNest} analysis in the form of corner plots. First we show the sensitivities on $\mu_{ij}$ parameters in~\cref{fig:Cornerplt-NO-mu} (\cref{fig:Cornerplt-IO-mu}) for the normal (inverted) mass ordering. We use flat priors for both phase and magnitude of $\mu_{ij}$ as well as for $\delta_{CP}$. We marginalize over $\sin^2\theta_{23}$ and fix other parameters at their central values as discussed in the main text. We then translate these to sensitivities on the $\eta_{\alpha \beta}$ parameters using the mapping discussed in subsection.~\ref{subsec:ParameterMapping}, setting $m_1 = 0.1$ eV ($m_3 = 0.1$ eV) and $\epsilon_{11}=0\ (\epsilon_{33}=0)$ for NO (IO) as benchmark. The obtained sensitivities in terms of $\eta_{\alpha\beta}$ is depicted in~\cref{fig:Cornerplt-NO} (\cref{fig:Cornerplt-IO}) for NO(IO). Within these figures, the diagonal subplots show marginal distributions as function of just one parameter while the off-diagonal subplots show the 2D marginalized plots with the dark (light) shaded regions showing the 68\% (95\%) confidence intervals for the expected sensitivities to oscillation and NSI parameters that DUNE experiment will be able to probe. 

\clearpage
\bibliographystyle{JHEP.bst}
\bibliography{scalarNSI.bib}

\providecommand{\href}[2]{#2}\begingroup\raggedright\begin{thebibliography}{10}

\bibitem{Mikheev:1986gs}
S.~Mikheyev and A.~Smirnov, {\it {Resonance Amplification of Oscillations in
  Matter and Spectroscopy of Solar Neutrinos}},  {\em Sov. J. Nucl. Phys.} {\bf
  42} (1985) 913--917.

\bibitem{Smirnov:2004zv}
A.~Smirnov, {\it {The MSW effect and matter effects in neutrino oscillations}},
   {\em Phys. Scripta T} {\bf 121} (2005) 57--64,
  [\href{http://arxiv.org/abs/hep-ph/0412391}{{\tt hep-ph/0412391}}].

\bibitem{Wolfenstein:1977ue}
L.~Wolfenstein, {\it {Neutrino Oscillations in Matter}},  {\em Phys. Rev. D}
  {\bf 17} (1978) 2369--2374. [,294(1977)].

\bibitem{SNO:2002tuh}
{\bf SNO} Collaboration, Q.~R. Ahmad et~al., {\it {Direct evidence for neutrino
  flavor transformation from neutral current interactions in the Sudbury
  Neutrino Observatory}},  {\em Phys. Rev. Lett.} {\bf 89} (2002) 011301,
  [\href{http://arxiv.org/abs/nucl-ex/0204008}{{\tt nucl-ex/0204008}}].

\bibitem{KamLAND:2002uet}
{\bf KamLAND} Collaboration, K.~Eguchi et~al., {\it {First results from
  KamLAND: Evidence for reactor anti-neutrino disappearance}},  {\em Phys. Rev.
  Lett.} {\bf 90} (2003) 021802,
  [\href{http://arxiv.org/abs/hep-ex/0212021}{{\tt hep-ex/0212021}}].

\bibitem{T2K:2011ypd}
{\bf T2K} Collaboration, K.~Abe et~al., {\it {Indication of Electron Neutrino
  Appearance from an Accelerator-produced Off-axis Muon Neutrino Beam}},  {\em
  Phys. Rev. Lett.} {\bf 107} (2011) 041801,
  [\href{http://arxiv.org/abs/1106.2822}{{\tt arXiv:1106.2822}}].

\bibitem{DoubleChooz:2011ymz}
{\bf Double Chooz} Collaboration, Y.~Abe et~al., {\it {Indication of Reactor
  $\bar{\nu}_e$ Disappearance in the Double Chooz Experiment}},  {\em Phys.
  Rev. Lett.} {\bf 108} (2012) 131801,
  [\href{http://arxiv.org/abs/1112.6353}{{\tt arXiv:1112.6353}}].

\bibitem{DayaBay:2012fng}
{\bf Daya Bay} Collaboration, F.~P. An et~al., {\it {Observation of
  electron-antineutrino disappearance at Daya Bay}},  {\em Phys. Rev. Lett.}
  {\bf 108} (2012) 171803, [\href{http://arxiv.org/abs/1203.1669}{{\tt
  arXiv:1203.1669}}].

\bibitem{RENO:2012mkc}
{\bf RENO} Collaboration, J.~K. Ahn et~al., {\it {Observation of Reactor
  Electron Antineutrino Disappearance in the RENO Experiment}},  {\em Phys.
  Rev. Lett.} {\bf 108} (2012) 191802,
  [\href{http://arxiv.org/abs/1204.0626}{{\tt arXiv:1204.0626}}].

\bibitem{T2K:2011qtm}
{\bf T2K} Collaboration, K.~Abe et~al., {\it {The T2K Experiment}},  {\em Nucl.
  Instrum. Meth. A} {\bf 659} (2011) 106--135,
  [\href{http://arxiv.org/abs/1106.1238}{{\tt arXiv:1106.1238}}].

\bibitem{NOvA:2007rmc}
{\bf NOvA} Collaboration, D.~S. Ayres et~al., {\it {The NOvA Technical Design
  Report}}, .

\bibitem{IceCube:2006tjp}
{\bf IceCube} Collaboration, A.~Achterberg et~al., {\it {First Year Performance
  of The IceCube Neutrino Telescope}},  {\em Astropart. Phys.} {\bf 26} (2006)
  155--173, [\href{http://arxiv.org/abs/astro-ph/0604450}{{\tt
  astro-ph/0604450}}].

\bibitem{Super-Kamiokande:2004orf}
{\bf Super-Kamiokande} Collaboration, Y.~Ashie et~al., {\it {Evidence for an
  oscillatory signature in atmospheric neutrino oscillation}},  {\em Phys. Rev.
  Lett.} {\bf 93} (2004) 101801,
  [\href{http://arxiv.org/abs/hep-ex/0404034}{{\tt hep-ex/0404034}}].

\bibitem{Gavela:2008ra}
M.~B. Gavela, D.~Hernandez, T.~Ota, and W.~Winter, {\it {Large gauge invariant
  non-standard neutrino interactions}},  {\em Phys. Rev. D} {\bf 79} (2009)
  013007, [\href{http://arxiv.org/abs/0809.3451}{{\tt arXiv:0809.3451}}].

\bibitem{Antusch:2008tz}
S.~Antusch, J.~P. Baumann, and E.~Fernandez-Martinez, {\it {Non-Standard
  Neutrino Interactions with Matter from Physics Beyond the Standard Model}},
  {\em Nucl. Phys. B} {\bf 810} (2009) 369--388,
  [\href{http://arxiv.org/abs/0807.1003}{{\tt arXiv:0807.1003}}].

\bibitem{Ohlsson:2012kf}
T.~Ohlsson, {\it {Status of non-standard neutrino interactions}},  {\em Rept.
  Prog. Phys.} {\bf 76} (2013) 044201,
  [\href{http://arxiv.org/abs/1209.2710}{{\tt arXiv:1209.2710}}].

\bibitem{Proceedings:2019qno}
{\em {Neutrino Non-Standard Interactions: A Status Report}}, vol.~2, 2019.

\bibitem{AristizabalSierra:2018eqm}
D.~Aristizabal~Sierra, V.~De~Romeri, and N.~Rojas, {\it {COHERENT analysis of
  neutrino generalized interactions}},  {\em Phys. Rev. D} {\bf 98} (2018)
  075018, [\href{http://arxiv.org/abs/1806.07424}{{\tt arXiv:1806.07424}}].

\bibitem{Giunti:2019xpr}
C.~Giunti, {\it {General COHERENT Constraints on Neutrino Non-Standard
  Interactions}},  \href{http://arxiv.org/abs/1909.00466}{{\tt
  arXiv:1909.00466}}.

\bibitem{Dutta:2022fdt}
B.~Dutta, S.~Ghosh, T.~Li, A.~Thompson, and A.~Verma, {\it {Non-standard
  neutrino interactions in light mediator models at reactor experiments}},
  {\em JHEP} {\bf 03} (2023) 163, [\href{http://arxiv.org/abs/2209.13566}{{\tt
  arXiv:2209.13566}}].

\bibitem{DUNE:2020ypp}
{\bf DUNE} Collaboration, B.~Abi et~al., {\it {Deep Underground Neutrino
  Experiment (DUNE), Far Detector Technical Design Report, Volume II: DUNE
  Physics}},  \href{http://arxiv.org/abs/2002.03005}{{\tt arXiv:2002.03005}}.

\bibitem{JUNO:2015zny}
{\bf JUNO} Collaboration, F.~An et~al., {\it {Neutrino Physics with JUNO}},
  {\em J. Phys. G} {\bf 43} (2016), no.~3 030401,
  [\href{http://arxiv.org/abs/1507.05613}{{\tt arXiv:1507.05613}}].

\bibitem{Ge:2018uhz}
S.-F. Ge and S.~J. Parke, {\it {Scalar Nonstandard Interactions in Neutrino
  Oscillation}},  {\em Phys. Rev. Lett.} {\bf 122} (2019), no.~21 211801,
  [\href{http://arxiv.org/abs/1812.08376}{{\tt arXiv:1812.08376}}].

\bibitem{Ge:2019tdi}
S.-F. Ge and H.~Murayama, {\it {Apparent CPT Violation in Neutrino Oscillation
  from Dark Non-Standard Interactions}},
  \href{http://arxiv.org/abs/1904.02518}{{\tt arXiv:1904.02518}}.

\bibitem{Denton:2022pxt}
P.~B. Denton, A.~Giarnetti, and D.~Meloni, {\it {How to identify different new
  neutrino oscillation physics scenarios at DUNE}},  {\em JHEP} {\bf 02} (2023)
  210, [\href{http://arxiv.org/abs/2210.00109}{{\tt arXiv:2210.00109}}].

\bibitem{Medhi:2021wxj}
A.~Medhi, D.~Dutta, and M.~M. Devi, {\it {Exploring the effects of scalar non
  standard interactions on the CP violation sensitivity at DUNE}},  {\em JHEP}
  {\bf 06} (2022) 129, [\href{http://arxiv.org/abs/2111.12943}{{\tt
  arXiv:2111.12943}}].

\bibitem{Medhi:2022qmu}
A.~Medhi, M.~M. Devi, and D.~Dutta, {\it {Imprints of scalar NSI on the
  CP-violation sensitivity using synergy among DUNE, T2HK and T2HKK}},  {\em
  JHEP} {\bf 01} (2023) 079, [\href{http://arxiv.org/abs/2209.05287}{{\tt
  arXiv:2209.05287}}].

\bibitem{Medhi:2023ebi}
A.~Medhi, A.~Sarker, and M.~M. Devi, {\it {Scalar NSI: A unique tool for
  constraining absolute neutrino masses via $\nu$-oscillations}},
  \href{http://arxiv.org/abs/2307.05348}{{\tt arXiv:2307.05348}}.

\bibitem{Singha:2023set}
D.~K. Singha, R.~Majhi, L.~Panda, M.~Ghosh, and R.~Mohanta, {\it {Study of
  Scalar Non Standard Interaction at Protvino to Super-ORCA experiment}},
  \href{http://arxiv.org/abs/2308.10789}{{\tt arXiv:2308.10789}}.

\bibitem{Sarker:2023qzp}
A.~Sarker, A.~Medhi, D.~Bezboruah, M.~M. Devi, and D.~Dutta, {\it {Impact of
  scalar NSI on the neutrino mass hierarchy sensitivity at DUNE, T2HK and
  T2HKK}},  \href{http://arxiv.org/abs/2309.12249}{{\tt arXiv:2309.12249}}.

\bibitem{Gupta:2023wct}
A.~Gupta, D.~Majumdar, and S.~Prakash, {\it {Neutrino oscillation measurements
  with JUNO in the presence of scalar NSI}},
  \href{http://arxiv.org/abs/2306.07343}{{\tt arXiv:2306.07343}}.

\bibitem{ESSnuSB:2023lbg}
{\bf ESSnuSB} Collaboration, J.~Aguilar et~al., {\it {Study of non-standard
  interaction mediated by a scalar field at ESSnuSB experiment}},
  \href{http://arxiv.org/abs/2310.10749}{{\tt arXiv:2310.10749}}.

\bibitem{T2K:2021xwb}
{\bf T2K} Collaboration, K.~Abe et~al., {\it {Improved constraints on neutrino
  mixing from the T2K experiment with $\mathbf{3.13\times10^{21}}$ protons on
  target}},  {\em Phys. Rev. D} {\bf 103} (2021), no.~11 112008,
  [\href{http://arxiv.org/abs/2101.03779}{{\tt arXiv:2101.03779}}].

\bibitem{T2K:2023smv}
{\bf T2K} Collaboration, K.~Abe et~al., {\it {Measurements of neutrino
  oscillation parameters from the T2K experiment using $3.6\times 10^{21}$
  protons on target}},  {\em Eur. Phys. J. C} {\bf 83} (2023), no.~9 782,
  [\href{http://arxiv.org/abs/2303.03222}{{\tt arXiv:2303.03222}}].

\bibitem{NOvA:2021nfi}
{\bf NOvA} Collaboration, M.~A. Acero et~al., {\it {Improved measurement of
  neutrino oscillation parameters by the NOvA experiment}},  {\em Phys. Rev. D}
  {\bf 106} (2022), no.~3 032004, [\href{http://arxiv.org/abs/2108.08219}{{\tt
  arXiv:2108.08219}}].

\bibitem{NOvA:2023iam}
{\bf NOvA, R. Group} Collaboration, M.~A. Acero et~al., {\it {Expanding
  neutrino oscillation parameter measurements in NOvA using a Bayesian
  approach}},  \href{http://arxiv.org/abs/2311.07835}{{\tt arXiv:2311.07835}}.

\bibitem{Kelly:2020fkv}
K.~J. Kelly, P.~A.~N. Machado, S.~J. Parke, Y.~F. Perez-Gonzalez, and R.~Z.
  Funchal, {\it {Neutrino mass ordering in light of recent data}},  {\em Phys.
  Rev. D} {\bf 103} (2021), no.~1 013004,
  [\href{http://arxiv.org/abs/2007.08526}{{\tt arXiv:2007.08526}}].

\bibitem{Esteban:2020cvm}
I.~Esteban, M.~C. Gonzalez-Garcia, M.~Maltoni, T.~Schwetz, and A.~Zhou, {\it
  {The fate of hints: updated global analysis of three-flavor neutrino
  oscillations}},  {\em JHEP} {\bf 09} (2020) 178,
  [\href{http://arxiv.org/abs/2007.14792}{{\tt arXiv:2007.14792}}].

\bibitem{deSalas:2020pgw}
P.~F. de~Salas, D.~V. Forero, S.~Gariazzo, P.~Mart\'\i{}nez-Mirav\'e, O.~Mena,
  C.~A. Ternes, M.~T\'ortola, and J.~W.~F. Valle, {\it {2020 global
  reassessment of the neutrino oscillation picture}},  {\em JHEP} {\bf 02}
  (2021) 071, [\href{http://arxiv.org/abs/2006.11237}{{\tt arXiv:2006.11237}}].

\bibitem{deGouvea:2015ndi}
A.~de~Gouv\^ea and K.~J. Kelly, {\it {Non-standard Neutrino Interactions at
  DUNE}},  {\em Nucl. Phys. B} {\bf 908} (2016) 318--335,
  [\href{http://arxiv.org/abs/1511.05562}{{\tt arXiv:1511.05562}}].

\bibitem{Coloma:2015kiu}
P.~Coloma, {\it {Non-Standard Interactions in propagation at the Deep
  Underground Neutrino Experiment}},  {\em JHEP} {\bf 03} (2016) 016,
  [\href{http://arxiv.org/abs/1511.06357}{{\tt arXiv:1511.06357}}].

\bibitem{Dutta:2020scq}
B.~Dutta, S.~Ghosh, and T.~Li, {\it {Explaining $(g-2)_{\mu,e}$, the KOTO
  anomaly and the MiniBooNE excess in an extended Higgs model with sterile
  neutrinos}},  {\em Phys. Rev. D} {\bf 102} (2020), no.~5 055017,
  [\href{http://arxiv.org/abs/2006.01319}{{\tt arXiv:2006.01319}}].

\bibitem{Berryman:2015nua}
J.~M. Berryman, A.~de~Gouv\^ea, K.~J. Kelly, and A.~Kobach, {\it {Sterile
  neutrino at the Deep Underground Neutrino Experiment}},  {\em Phys. Rev. D}
  {\bf 92} (2015), no.~7 073012, [\href{http://arxiv.org/abs/1507.03986}{{\tt
  arXiv:1507.03986}}].

\bibitem{Berryman:2016szd}
J.~M. Berryman, A.~de~Gouv\^ea, K.~J. Kelly, O.~L.~G. Peres, and Z.~Tabrizi,
  {\it {Large, Extra Dimensions at the Deep Underground Neutrino Experiment}},
  {\em Phys. Rev. D} {\bf 94} (2016), no.~3 033006,
  [\href{http://arxiv.org/abs/1603.00018}{{\tt arXiv:1603.00018}}].

\bibitem{deGouvea:2016pom}
A.~de~Gouv\^ea and K.~J. Kelly, {\it {False Signals of CP-Invariance Violation
  at DUNE}},  \href{http://arxiv.org/abs/1605.09376}{{\tt arXiv:1605.09376}}.

\bibitem{deGouvea:2017yvn}
A.~de~Gouv\^ea and K.~J. Kelly, {\it {Neutrino vs. Antineutrino Oscillation
  Parameters at DUNE and Hyper-Kamiokande}},  {\em Phys. Rev. D} {\bf 96}
  (2017), no.~9 095018, [\href{http://arxiv.org/abs/1709.06090}{{\tt
  arXiv:1709.06090}}].

\bibitem{Feroz:2008xx}
F.~Feroz, M.~P. Hobson, and M.~Bridges, {\it {MultiNest: an efficient and
  robust Bayesian inference tool for cosmology and particle physics}},  {\em
  Mon. Not. Roy. Astron. Soc.} {\bf 398} (2009) 1601--1614,
  [\href{http://arxiv.org/abs/0809.3437}{{\tt arXiv:0809.3437}}].

\bibitem{Buchner:2014nha}
J.~Buchner, A.~Georgakakis, K.~Nandra, L.~Hsu, C.~Rangel, M.~Brightman,
  A.~Merloni, M.~Salvato, J.~Donley, and D.~Kocevski, {\it {X-ray spectral
  modelling of the AGN obscuring region in the CDFS: Bayesian model selection
  and catalogue}},  {\em Astron. Astrophys.} {\bf 564} (2014) A125,
  [\href{http://arxiv.org/abs/1402.0004}{{\tt arXiv:1402.0004}}].

\bibitem{Damour:2010rm}
T.~Damour and J.~F. Donoghue, {\it {Phenomenology of the Equivalence Principle
  with Light Scalars}},  {\em Class. Quant. Grav.} {\bf 27} (2010) 202001,
  [\href{http://arxiv.org/abs/1007.2790}{{\tt arXiv:1007.2790}}].

\bibitem{Brzeminski:2022sde}
D.~Brzeminski, Z.~Chacko, A.~Dev, I.~Flood, and A.~Hook, {\it {Searching for a
  fifth force with atomic and nuclear clocks}},  {\em Phys. Rev. D} {\bf 106}
  (2022), no.~9 095031, [\href{http://arxiv.org/abs/2207.14310}{{\tt
  arXiv:2207.14310}}].

\bibitem{Schlamminger:2007ht}
S.~Schlamminger, K.~Y. Choi, T.~A. Wagner, J.~H. Gundlach, and E.~G.
  Adelberger, {\it {Test of the equivalence principle using a rotating torsion
  balance}},  {\em Phys. Rev. Lett.} {\bf 100} (2008) 041101,
  [\href{http://arxiv.org/abs/0712.0607}{{\tt arXiv:0712.0607}}].

\bibitem{Berge:2017ovy}
J.~Berg\'e, P.~Brax, G.~M\'etris, M.~Pernot-Borr\`as, P.~Touboul, and J.-P.
  Uzan, {\it {MICROSCOPE Mission: First Constraints on the Violation of the
  Weak Equivalence Principle by a Light Scalar Dilaton}},  {\em Phys. Rev.
  Lett.} {\bf 120} (2018), no.~14 141101,
  [\href{http://arxiv.org/abs/1712.00483}{{\tt arXiv:1712.00483}}].

\bibitem{MICROSCOPE:2022doy}
{\bf MICROSCOPE} Collaboration, P.~Touboul et~al., {\it {MICROSCOPE Mission:
  Final Results of the Test of the Equivalence Principle}},  {\em Phys. Rev.
  Lett.} {\bf 129} (2022), no.~12 121102,
  [\href{http://arxiv.org/abs/2209.15487}{{\tt arXiv:2209.15487}}].

\bibitem{Sandner:2023ptm}
S.~Sandner, M.~Escudero, and S.~J. Witte, {\it {Precision CMB constraints on
  eV-scale bosons coupled to neutrinos}},  {\em Eur. Phys. J. C} {\bf 83}
  (2023), no.~8 709, [\href{http://arxiv.org/abs/2305.01692}{{\tt
  arXiv:2305.01692}}].

\bibitem{NeffEscudero}
M.~Escudero, ``{Neutrino (self)-Interactions in Cosmology}.'' CERN Neutrino
  Platform Pheno Week, 2023.

\bibitem{Babu:2019iml}
K.~S. Babu, G.~Chauhan, and P.~S. Bhupal~Dev, {\it {Neutrino nonstandard
  interactions via light scalars in the Earth, Sun, supernovae, and the early
  Universe}},  {\em Phys. Rev. D} {\bf 101} (2020), no.~9 095029,
  [\href{http://arxiv.org/abs/1912.13488}{{\tt arXiv:1912.13488}}].

\bibitem{Hardy:2016kme}
E.~Hardy and R.~Lasenby, {\it {Stellar cooling bounds on new light particles:
  plasma mixing effects}},  {\em JHEP} {\bf 02} (2017) 033,
  [\href{http://arxiv.org/abs/1611.05852}{{\tt arXiv:1611.05852}}].

\bibitem{Borexino:2017rsf}
{\bf Borexino} Collaboration, M.~Agostini et~al., {\it {First Simultaneous
  Precision Spectroscopy of $pp$, $^7$Be, and $pep$ Solar Neutrinos with
  Borexino Phase-II}},  {\em Phys. Rev. D} {\bf 100} (2019), no.~8 082004,
  [\href{http://arxiv.org/abs/1707.09279}{{\tt arXiv:1707.09279}}].

\bibitem{Smirnov:2019cae}
A.~Y. Smirnov and X.-J. Xu, {\it {Wolfenstein potentials for neutrinos induced
  by ultra-light mediators}},  {\em JHEP} {\bf 12} (2019) 046,
  [\href{http://arxiv.org/abs/1909.07505}{{\tt arXiv:1909.07505}}].

\bibitem{TEXONO:2009knm}
{\bf TEXONO} Collaboration, M.~Deniz et~al., {\it {Measurement of Nu(e)-bar
  -Electron Scattering Cross-Section with a CsI(Tl) Scintillating Crystal Array
  at the Kuo-Sheng Nuclear Power Reactor}},  {\em Phys. Rev. D} {\bf 81} (2010)
  072001, [\href{http://arxiv.org/abs/0911.1597}{{\tt arXiv:0911.1597}}].

\bibitem{BOREXINO:2018ohr}
{\bf BOREXINO} Collaboration, M.~Agostini et~al., {\it {Comprehensive
  measurement of $pp$-chain solar neutrinos}},  {\em Nature} {\bf 562} (2018),
  no.~7728 505--510.

\bibitem{Beda:2013mta}
A.~G. Beda, V.~B. Brudanin, V.~G. Egorov, D.~V. Medvedev, V.~S. Pogosov, E.~A.
  Shevchik, M.~V. Shirchenko, A.~S. Starostin, and I.~V. Zhitnikov, {\it {Gemma
  experiment: The results of neutrino magnetic moment search}},  {\em Phys.
  Part. Nucl. Lett.} {\bf 10} (2013) 139--143.

\bibitem{XENON:2022ltv}
{\bf XENON} Collaboration, E.~Aprile et~al., {\it {Search for New Physics in
  Electronic Recoil Data from XENONnT}},  {\em Phys. Rev. Lett.} {\bf 129}
  (2022), no.~16 161805, [\href{http://arxiv.org/abs/2207.11330}{{\tt
  arXiv:2207.11330}}].

\bibitem{LZ:2022lsv}
{\bf LZ} Collaboration, J.~Aalbers et~al., {\it {First Dark Matter Search
  Results from the LUX-ZEPLIN (LZ) Experiment}},  {\em Phys. Rev. Lett.} {\bf
  131} (2023), no.~4 041002, [\href{http://arxiv.org/abs/2207.03764}{{\tt
  arXiv:2207.03764}}].

\bibitem{Akimov:2017ade}
{\bf COHERENT} Collaboration, D.~Akimov et~al., {\it {Observation of Coherent
  Elastic Neutrino-Nucleus Scattering}},  {\em Science} {\bf 357} (2017),
  no.~6356 1123--1126, [\href{http://arxiv.org/abs/1708.01294}{{\tt
  arXiv:1708.01294}}].

\bibitem{Akimov:2018vzs}
{\bf COHERENT} Collaboration, D.~Akimov et~al., {\it {COHERENT Collaboration
  data release from the first observation of coherent elastic neutrino-nucleus
  scattering}},  \href{http://arxiv.org/abs/1804.09459}{{\tt
  arXiv:1804.09459}}.

\bibitem{Akimov:2018ghi}
{\bf COHERENT} Collaboration, D.~Akimov et~al., {\it {COHERENT 2018 at the
  Spallation Neutron Source}},  \href{http://arxiv.org/abs/1803.09183}{{\tt
  arXiv:1803.09183}}.

\bibitem{Akimov:2019xdj}
{\bf COHERENT} Collaboration, D.~Akimov et~al., {\it {Sensitivity of the
  COHERENT Experiment to Accelerator-Produced Dark Matter}},
  \href{http://arxiv.org/abs/1911.06422}{{\tt arXiv:1911.06422}}.

\bibitem{Akimov:2020pdx}
{\bf COHERENT} Collaboration, D.~Akimov et~al., {\it {First Detection of
  Coherent Elastic Neutrino-Nucleus Scattering on Argon}},
  \href{http://arxiv.org/abs/2003.10630}{{\tt arXiv:2003.10630}}.

\bibitem{COHERENT:2021pvd}
{\bf COHERENT} Collaboration, D.~Akimov et~al., {\it {First Probe of Sub-GeV
  Dark Matter Beyond the Cosmological Expectation with the COHERENT CsI
  Detector at the SNS}},  \href{http://arxiv.org/abs/2110.11453}{{\tt
  arXiv:2110.11453}}.

\end{thebibliography}\endgroup

\end{document}